\documentclass[usenatbib,usegraphicx]{mn2e}
\usepackage{amsfonts}
\usepackage{amsmath}
\usepackage{amssymb}
\usepackage{comment}
\usepackage{color}
\definecolor{AliceBlue}{rgb}{0.94,0.97,1.00}
\definecolor{AntiqueWhite1}{rgb}{1.00,0.94,0.86}
\definecolor{AntiqueWhite2}{rgb}{0.93,0.87,0.80}
\definecolor{AntiqueWhite3}{rgb}{0.80,0.75,0.69}
\definecolor{AntiqueWhite4}{rgb}{0.55,0.51,0.47}
\definecolor{AntiqueWhite}{rgb}{0.98,0.92,0.84}
\definecolor{BlanchedAlmond}{rgb}{1.00,0.92,0.80}
\definecolor{BlueViolet}{rgb}{0.54,0.17,0.89}
\definecolor{CadetBlue1}{rgb}{0.60,0.96,1.00}
\definecolor{CadetBlue2}{rgb}{0.56,0.90,0.93}
\definecolor{CadetBlue3}{rgb}{0.48,0.77,0.80}
\definecolor{CadetBlue4}{rgb}{0.33,0.53,0.55}
\definecolor{CadetBlue}{rgb}{0.37,0.62,0.63}
\definecolor{CornflowerBlue}{rgb}{0.39,0.58,0.93}
\definecolor{DarkBlue}{rgb}{0.00,0.00,0.55}
\definecolor{DarkCyan}{rgb}{0.00,0.55,0.55}
\definecolor{DarkGoldenrod1}{rgb}{1.00,0.73,0.06}
\definecolor{DarkGoldenrod2}{rgb}{0.93,0.68,0.05}
\definecolor{DarkGoldenrod3}{rgb}{0.80,0.58,0.05}
\definecolor{DarkGoldenrod4}{rgb}{0.55,0.40,0.03}
\definecolor{DarkGoldenrod}{rgb}{0.72,0.53,0.04}
\definecolor{DarkGray}{rgb}{0.66,0.66,0.66}
\definecolor{DarkGreen}{rgb}{0.00,0.39,0.00}
\definecolor{DarkGrey}{rgb}{0.66,0.66,0.66}
\definecolor{DarkKhaki}{rgb}{0.74,0.72,0.42}
\definecolor{DarkMagenta}{rgb}{0.55,0.00,0.55}
\definecolor{DarkOliveGreen1}{rgb}{0.79,1.00,0.44}
\definecolor{DarkOliveGreen2}{rgb}{0.74,0.93,0.41}
\definecolor{DarkOliveGreen3}{rgb}{0.64,0.80,0.35}
\definecolor{DarkOliveGreen4}{rgb}{0.43,0.55,0.24}
\definecolor{DarkOliveGreen}{rgb}{0.33,0.42,0.18}
\definecolor{DarkOrange1}{rgb}{1.00,0.50,0.00}
\definecolor{DarkOrange2}{rgb}{0.93,0.46,0.00}
\definecolor{DarkOrange3}{rgb}{0.80,0.40,0.00}
\definecolor{DarkOrange4}{rgb}{0.55,0.27,0.00}
\definecolor{DarkOrange}{rgb}{1.00,0.55,0.00}
\definecolor{DarkOrchid1}{rgb}{0.75,0.24,1.00}
\definecolor{DarkOrchid2}{rgb}{0.70,0.23,0.93}
\definecolor{DarkOrchid3}{rgb}{0.60,0.20,0.80}
\definecolor{DarkOrchid4}{rgb}{0.41,0.13,0.55}
\definecolor{DarkOrchid}{rgb}{0.60,0.20,0.80}
\definecolor{DarkRed}{rgb}{0.55,0.00,0.00}
\definecolor{DarkSalmon}{rgb}{0.91,0.59,0.48}
\definecolor{DarkSeaGreen1}{rgb}{0.76,1.00,0.76}
\definecolor{DarkSeaGreen2}{rgb}{0.71,0.93,0.71}
\definecolor{DarkSeaGreen3}{rgb}{0.61,0.80,0.61}
\definecolor{DarkSeaGreen4}{rgb}{0.41,0.55,0.41}
\definecolor{DarkSeaGreen}{rgb}{0.56,0.74,0.56}
\definecolor{DarkSlateBlue}{rgb}{0.28,0.24,0.55}
\definecolor{DarkSlateGray1}{rgb}{0.59,1.00,1.00}
\definecolor{DarkSlateGray2}{rgb}{0.55,0.93,0.93}
\definecolor{DarkSlateGray3}{rgb}{0.47,0.80,0.80}
\definecolor{DarkSlateGray4}{rgb}{0.32,0.55,0.55}
\definecolor{DarkSlateGray}{rgb}{0.18,0.31,0.31}
\definecolor{DarkSlateGrey}{rgb}{0.18,0.31,0.31}
\definecolor{DarkTurquoise}{rgb}{0.00,0.81,0.82}
\definecolor{DarkViolet}{rgb}{0.58,0.00,0.83}
\definecolor{DeepPink1}{rgb}{1.00,0.08,0.58}
\definecolor{DeepPink2}{rgb}{0.93,0.07,0.54}
\definecolor{DeepPink3}{rgb}{0.80,0.06,0.46}
\definecolor{DeepPink4}{rgb}{0.55,0.04,0.31}
\definecolor{DeepPink}{rgb}{1.00,0.08,0.58}
\definecolor{DeepSkyBlue1}{rgb}{0.00,0.75,1.00}
\definecolor{DeepSkyBlue2}{rgb}{0.00,0.70,0.93}
\definecolor{DeepSkyBlue3}{rgb}{0.00,0.60,0.80}
\definecolor{DeepSkyBlue4}{rgb}{0.00,0.41,0.55}
\definecolor{DeepSkyBlue}{rgb}{0.00,0.75,1.00}
\definecolor{DimGray}{rgb}{0.41,0.41,0.41}
\definecolor{DimGrey}{rgb}{0.41,0.41,0.41}
\definecolor{DodgerBlue1}{rgb}{0.12,0.56,1.00}
\definecolor{DodgerBlue2}{rgb}{0.11,0.53,0.93}
\definecolor{DodgerBlue3}{rgb}{0.09,0.45,0.80}
\definecolor{DodgerBlue4}{rgb}{0.06,0.31,0.55}
\definecolor{DodgerBlue}{rgb}{0.12,0.56,1.00}
\definecolor{FloralWhite}{rgb}{1.00,0.98,0.94}
\definecolor{ForestGreen}{rgb}{0.13,0.55,0.13}
\definecolor{GhostWhite}{rgb}{0.97,0.97,1.00}
\definecolor{GreenYellow}{rgb}{0.68,1.00,0.18}
\definecolor{HotPink1}{rgb}{1.00,0.43,0.71}
\definecolor{HotPink2}{rgb}{0.93,0.42,0.65}
\definecolor{HotPink3}{rgb}{0.80,0.38,0.56}
\definecolor{HotPink4}{rgb}{0.55,0.23,0.38}
\definecolor{HotPink}{rgb}{1.00,0.41,0.71}
\definecolor{IndianRed1}{rgb}{1.00,0.42,0.42}
\definecolor{IndianRed2}{rgb}{0.93,0.39,0.39}
\definecolor{IndianRed3}{rgb}{0.80,0.33,0.33}
\definecolor{IndianRed4}{rgb}{0.55,0.23,0.23}
\definecolor{IndianRed}{rgb}{0.80,0.36,0.36}
\definecolor{LavenderBlush1}{rgb}{1.00,0.94,0.96}
\definecolor{LavenderBlush2}{rgb}{0.93,0.88,0.90}
\definecolor{LavenderBlush3}{rgb}{0.80,0.76,0.77}
\definecolor{LavenderBlush4}{rgb}{0.55,0.51,0.53}
\definecolor{LavenderBlush}{rgb}{1.00,0.94,0.96}
\definecolor{LawnGreen}{rgb}{0.49,0.99,0.00}
\definecolor{LemonChiffon1}{rgb}{1.00,0.98,0.80}
\definecolor{LemonChiffon2}{rgb}{0.93,0.91,0.75}
\definecolor{LemonChiffon3}{rgb}{0.80,0.79,0.65}
\definecolor{LemonChiffon4}{rgb}{0.55,0.54,0.44}
\definecolor{LemonChiffon}{rgb}{1.00,0.98,0.80}
\definecolor{LightBlue1}{rgb}{0.75,0.94,1.00}
\definecolor{LightBlue2}{rgb}{0.70,0.87,0.93}
\definecolor{LightBlue3}{rgb}{0.60,0.75,0.80}
\definecolor{LightBlue4}{rgb}{0.41,0.51,0.55}
\definecolor{LightBlue}{rgb}{0.68,0.85,0.90}
\definecolor{LightCoral}{rgb}{0.94,0.50,0.50}
\definecolor{LightCyan1}{rgb}{0.88,1.00,1.00}
\definecolor{LightCyan2}{rgb}{0.82,0.93,0.93}
\definecolor{LightCyan3}{rgb}{0.71,0.80,0.80}
\definecolor{LightCyan4}{rgb}{0.48,0.55,0.55}
\definecolor{LightCyan}{rgb}{0.88,1.00,1.00}
\definecolor{LightGoldenrod1}{rgb}{1.00,0.93,0.55}
\definecolor{LightGoldenrod2}{rgb}{0.93,0.86,0.51}
\definecolor{LightGoldenrod3}{rgb}{0.80,0.75,0.44}
\definecolor{LightGoldenrod4}{rgb}{0.55,0.51,0.30}
\definecolor{LightGoldenrodYellow}{rgb}{0.98,0.98,0.82}
\definecolor{LightGoldenrod}{rgb}{0.93,0.87,0.51}
\definecolor{LightGray}{rgb}{0.83,0.83,0.83}
\definecolor{LightGreen}{rgb}{0.56,0.93,0.56}
\definecolor{LightGrey}{rgb}{0.83,0.83,0.83}
\definecolor{LightPink1}{rgb}{1.00,0.68,0.73}
\definecolor{LightPink2}{rgb}{0.93,0.64,0.68}
\definecolor{LightPink3}{rgb}{0.80,0.55,0.58}
\definecolor{LightPink4}{rgb}{0.55,0.37,0.40}
\definecolor{LightPink}{rgb}{1.00,0.71,0.76}
\definecolor{LightSalmon1}{rgb}{1.00,0.63,0.48}
\definecolor{LightSalmon2}{rgb}{0.93,0.58,0.45}
\definecolor{LightSalmon3}{rgb}{0.80,0.51,0.38}
\definecolor{LightSalmon4}{rgb}{0.55,0.34,0.26}
\definecolor{LightSalmon}{rgb}{1.00,0.63,0.48}
\definecolor{LightSeaGreen}{rgb}{0.13,0.70,0.67}
\definecolor{LightSkyBlue1}{rgb}{0.69,0.89,1.00}
\definecolor{LightSkyBlue2}{rgb}{0.64,0.83,0.93}
\definecolor{LightSkyBlue3}{rgb}{0.55,0.71,0.80}
\definecolor{LightSkyBlue4}{rgb}{0.38,0.48,0.55}
\definecolor{LightSkyBlue}{rgb}{0.53,0.81,0.98}
\definecolor{LightSlateBlue}{rgb}{0.52,0.44,1.00}
\definecolor{LightSlateGray}{rgb}{0.47,0.53,0.60}
\definecolor{LightSlateGrey}{rgb}{0.47,0.53,0.60}
\definecolor{LightSteelBlue1}{rgb}{0.79,0.88,1.00}
\definecolor{LightSteelBlue2}{rgb}{0.74,0.82,0.93}
\definecolor{LightSteelBlue3}{rgb}{0.64,0.71,0.80}
\definecolor{LightSteelBlue4}{rgb}{0.43,0.48,0.55}
\definecolor{LightSteelBlue}{rgb}{0.69,0.77,0.87}
\definecolor{LightYellow1}{rgb}{1.00,1.00,0.88}
\definecolor{LightYellow2}{rgb}{0.93,0.93,0.82}
\definecolor{LightYellow3}{rgb}{0.80,0.80,0.71}
\definecolor{LightYellow4}{rgb}{0.55,0.55,0.48}
\definecolor{LightYellow}{rgb}{1.00,1.00,0.88}
\definecolor{LimeGreen}{rgb}{0.20,0.80,0.20}
\definecolor{MediumAquamarine}{rgb}{0.40,0.80,0.67}
\definecolor{MediumBlue}{rgb}{0.00,0.00,0.80}
\definecolor{MediumOrchid1}{rgb}{0.88,0.40,1.00}
\definecolor{MediumOrchid2}{rgb}{0.82,0.37,0.93}
\definecolor{MediumOrchid3}{rgb}{0.71,0.32,0.80}
\definecolor{MediumOrchid4}{rgb}{0.48,0.22,0.55}
\definecolor{MediumOrchid}{rgb}{0.73,0.33,0.83}
\definecolor{MediumPurple1}{rgb}{0.67,0.51,1.00}
\definecolor{MediumPurple2}{rgb}{0.62,0.47,0.93}
\definecolor{MediumPurple3}{rgb}{0.54,0.41,0.80}
\definecolor{MediumPurple4}{rgb}{0.36,0.28,0.55}
\definecolor{MediumPurple}{rgb}{0.58,0.44,0.86}
\definecolor{MediumSeaGreen}{rgb}{0.24,0.70,0.44}
\definecolor{MediumSlateBlue}{rgb}{0.48,0.41,0.93}
\definecolor{MediumSpringGreen}{rgb}{0.00,0.98,0.60}
\definecolor{MediumTurquoise}{rgb}{0.28,0.82,0.80}
\definecolor{MediumVioletRed}{rgb}{0.78,0.08,0.52}
\definecolor{MidnightBlue}{rgb}{0.10,0.10,0.44}
\definecolor{MintCream}{rgb}{0.96,1.00,0.98}
\definecolor{MistyRose1}{rgb}{1.00,0.89,0.88}
\definecolor{MistyRose2}{rgb}{0.93,0.84,0.82}
\definecolor{MistyRose3}{rgb}{0.80,0.72,0.71}
\definecolor{MistyRose4}{rgb}{0.55,0.49,0.48}
\definecolor{MistyRose}{rgb}{1.00,0.89,0.88}
\definecolor{NavajoWhite1}{rgb}{1.00,0.87,0.68}
\definecolor{NavajoWhite2}{rgb}{0.93,0.81,0.63}
\definecolor{NavajoWhite3}{rgb}{0.80,0.70,0.55}
\definecolor{NavajoWhite4}{rgb}{0.55,0.47,0.37}
\definecolor{NavajoWhite}{rgb}{1.00,0.87,0.68}
\definecolor{NavyBlue}{rgb}{0.00,0.00,0.50}
\definecolor{OldLace}{rgb}{0.99,0.96,0.90}
\definecolor{OliveDrab1}{rgb}{0.75,1.00,0.24}
\definecolor{OliveDrab2}{rgb}{0.70,0.93,0.23}
\definecolor{OliveDrab3}{rgb}{0.60,0.80,0.20}
\definecolor{OliveDrab4}{rgb}{0.41,0.55,0.13}
\definecolor{OliveDrab}{rgb}{0.42,0.56,0.14}
\definecolor{OrangeRed1}{rgb}{1.00,0.27,0.00}
\definecolor{OrangeRed2}{rgb}{0.93,0.25,0.00}
\definecolor{OrangeRed3}{rgb}{0.80,0.22,0.00}
\definecolor{OrangeRed4}{rgb}{0.55,0.15,0.00}
\definecolor{OrangeRed}{rgb}{1.00,0.27,0.00}
\definecolor{PaleGoldenrod}{rgb}{0.93,0.91,0.67}
\definecolor{PaleGreen1}{rgb}{0.60,1.00,0.60}
\definecolor{PaleGreen2}{rgb}{0.56,0.93,0.56}
\definecolor{PaleGreen3}{rgb}{0.49,0.80,0.49}
\definecolor{PaleGreen4}{rgb}{0.33,0.55,0.33}
\definecolor{PaleGreen}{rgb}{0.60,0.98,0.60}
\definecolor{PaleTurquoise1}{rgb}{0.73,1.00,1.00}
\definecolor{PaleTurquoise2}{rgb}{0.68,0.93,0.93}
\definecolor{PaleTurquoise3}{rgb}{0.59,0.80,0.80}
\definecolor{PaleTurquoise4}{rgb}{0.40,0.55,0.55}
\definecolor{PaleTurquoise}{rgb}{0.69,0.93,0.93}
\definecolor{PaleVioletRed1}{rgb}{1.00,0.51,0.67}
\definecolor{PaleVioletRed2}{rgb}{0.93,0.47,0.62}
\definecolor{PaleVioletRed3}{rgb}{0.80,0.41,0.54}
\definecolor{PaleVioletRed4}{rgb}{0.55,0.28,0.36}
\definecolor{PaleVioletRed}{rgb}{0.86,0.44,0.58}
\definecolor{PapayaWhip}{rgb}{1.00,0.94,0.84}
\definecolor{PeachPuff1}{rgb}{1.00,0.85,0.73}
\definecolor{PeachPuff2}{rgb}{0.93,0.80,0.68}
\definecolor{PeachPuff3}{rgb}{0.80,0.69,0.58}
\definecolor{PeachPuff4}{rgb}{0.55,0.47,0.40}
\definecolor{PeachPuff}{rgb}{1.00,0.85,0.73}
\definecolor{PowderBlue}{rgb}{0.69,0.88,0.90}
\definecolor{RosyBrown1}{rgb}{1.00,0.76,0.76}
\definecolor{RosyBrown2}{rgb}{0.93,0.71,0.71}
\definecolor{RosyBrown3}{rgb}{0.80,0.61,0.61}
\definecolor{RosyBrown4}{rgb}{0.55,0.41,0.41}
\definecolor{RosyBrown}{rgb}{0.74,0.56,0.56}
\definecolor{RoyalBlue1}{rgb}{0.28,0.46,1.00}
\definecolor{RoyalBlue2}{rgb}{0.26,0.43,0.93}
\definecolor{RoyalBlue3}{rgb}{0.23,0.37,0.80}
\definecolor{RoyalBlue4}{rgb}{0.15,0.25,0.55}
\definecolor{RoyalBlue}{rgb}{0.25,0.41,0.88}
\definecolor{SaddleBrown}{rgb}{0.55,0.27,0.07}
\definecolor{SandyBrown}{rgb}{0.96,0.64,0.38}
\definecolor{SeaGreen1}{rgb}{0.33,1.00,0.62}
\definecolor{SeaGreen2}{rgb}{0.31,0.93,0.58}
\definecolor{SeaGreen3}{rgb}{0.26,0.80,0.50}
\definecolor{SeaGreen4}{rgb}{0.18,0.55,0.34}
\definecolor{SeaGreen}{rgb}{0.18,0.55,0.34}
\definecolor{SkyBlue1}{rgb}{0.53,0.81,1.00}
\definecolor{SkyBlue2}{rgb}{0.49,0.75,0.93}
\definecolor{SkyBlue3}{rgb}{0.42,0.65,0.80}
\definecolor{SkyBlue4}{rgb}{0.29,0.44,0.55}
\definecolor{SkyBlue}{rgb}{0.53,0.81,0.92}
\definecolor{SlateBlue1}{rgb}{0.51,0.44,1.00}
\definecolor{SlateBlue2}{rgb}{0.48,0.40,0.93}
\definecolor{SlateBlue3}{rgb}{0.41,0.35,0.80}
\definecolor{SlateBlue4}{rgb}{0.28,0.24,0.55}
\definecolor{SlateBlue}{rgb}{0.42,0.35,0.80}
\definecolor{SlateGray1}{rgb}{0.78,0.89,1.00}
\definecolor{SlateGray2}{rgb}{0.73,0.83,0.93}
\definecolor{SlateGray3}{rgb}{0.62,0.71,0.80}
\definecolor{SlateGray4}{rgb}{0.42,0.48,0.55}
\definecolor{SlateGray}{rgb}{0.44,0.50,0.56}
\definecolor{SlateGrey}{rgb}{0.44,0.50,0.56}
\definecolor{SpringGreen1}{rgb}{0.00,1.00,0.50}
\definecolor{SpringGreen2}{rgb}{0.00,0.93,0.46}
\definecolor{SpringGreen3}{rgb}{0.00,0.80,0.40}
\definecolor{SpringGreen4}{rgb}{0.00,0.55,0.27}
\definecolor{SpringGreen}{rgb}{0.00,1.00,0.50}
\definecolor{SteelBlue1}{rgb}{0.39,0.72,1.00}
\definecolor{SteelBlue2}{rgb}{0.36,0.67,0.93}
\definecolor{SteelBlue3}{rgb}{0.31,0.58,0.80}
\definecolor{SteelBlue4}{rgb}{0.21,0.39,0.55}
\definecolor{SteelBlue}{rgb}{0.27,0.51,0.71}
\definecolor{VioletRed1}{rgb}{1.00,0.24,0.59}
\definecolor{VioletRed2}{rgb}{0.93,0.23,0.55}
\definecolor{VioletRed3}{rgb}{0.80,0.20,0.47}
\definecolor{VioletRed4}{rgb}{0.55,0.13,0.32}
\definecolor{VioletRed}{rgb}{0.82,0.13,0.56}
\definecolor{WhiteSmoke}{rgb}{0.96,0.96,0.96}
\definecolor{YellowGreen}{rgb}{0.60,0.80,0.20}
\definecolor{aliceblue}{rgb}{0.94,0.97,1.00}
\definecolor{antiquewhite}{rgb}{0.98,0.92,0.84}
\definecolor{aquamarine1}{rgb}{0.50,1.00,0.83}
\definecolor{aquamarine2}{rgb}{0.46,0.93,0.78}
\definecolor{aquamarine3}{rgb}{0.40,0.80,0.67}
\definecolor{aquamarine4}{rgb}{0.27,0.55,0.45}
\definecolor{aquamarine}{rgb}{0.50,1.00,0.83}
\definecolor{azure1}{rgb}{0.94,1.00,1.00}
\definecolor{azure2}{rgb}{0.88,0.93,0.93}
\definecolor{azure3}{rgb}{0.76,0.80,0.80}
\definecolor{azure4}{rgb}{0.51,0.55,0.55}
\definecolor{azure}{rgb}{0.94,1.00,1.00}
\definecolor{beige}{rgb}{0.96,0.96,0.86}
\definecolor{bisque1}{rgb}{1.00,0.89,0.77}
\definecolor{bisque2}{rgb}{0.93,0.84,0.72}
\definecolor{bisque3}{rgb}{0.80,0.72,0.62}
\definecolor{bisque4}{rgb}{0.55,0.49,0.42}
\definecolor{bisque}{rgb}{1.00,0.89,0.77}
\definecolor{black}{rgb}{0.00,0.00,0.00}
\definecolor{blanchedalmond}{rgb}{1.00,0.92,0.80}
\definecolor{blue1}{rgb}{0.00,0.00,1.00}
\definecolor{blue2}{rgb}{0.00,0.00,0.93}
\definecolor{blue3}{rgb}{0.00,0.00,0.80}
\definecolor{blue4}{rgb}{0.00,0.00,0.55}
\definecolor{blueviolet}{rgb}{0.54,0.17,0.89}
\definecolor{blue}{rgb}{0.00,0.00,1.00}
\definecolor{brown1}{rgb}{1.00,0.25,0.25}
\definecolor{brown2}{rgb}{0.93,0.23,0.23}
\definecolor{brown3}{rgb}{0.80,0.20,0.20}
\definecolor{brown4}{rgb}{0.55,0.14,0.14}
\definecolor{brown}{rgb}{0.65,0.16,0.16}
\definecolor{burlywood1}{rgb}{1.00,0.83,0.61}
\definecolor{burlywood2}{rgb}{0.93,0.77,0.57}
\definecolor{burlywood3}{rgb}{0.80,0.67,0.49}
\definecolor{burlywood4}{rgb}{0.55,0.45,0.33}
\definecolor{burlywood}{rgb}{0.87,0.72,0.53}
\definecolor{cadetblue}{rgb}{0.37,0.62,0.63}
\definecolor{chartreuse1}{rgb}{0.50,1.00,0.00}
\definecolor{chartreuse2}{rgb}{0.46,0.93,0.00}
\definecolor{chartreuse3}{rgb}{0.40,0.80,0.00}
\definecolor{chartreuse4}{rgb}{0.27,0.55,0.00}
\definecolor{chartreuse}{rgb}{0.50,1.00,0.00}
\definecolor{chocolate1}{rgb}{1.00,0.50,0.14}
\definecolor{chocolate2}{rgb}{0.93,0.46,0.13}
\definecolor{chocolate3}{rgb}{0.80,0.40,0.11}
\definecolor{chocolate4}{rgb}{0.55,0.27,0.07}
\definecolor{chocolate}{rgb}{0.82,0.41,0.12}
\definecolor{coral1}{rgb}{1.00,0.45,0.34}
\definecolor{coral2}{rgb}{0.93,0.42,0.31}
\definecolor{coral3}{rgb}{0.80,0.36,0.27}
\definecolor{coral4}{rgb}{0.55,0.24,0.18}
\definecolor{coral}{rgb}{1.00,0.50,0.31}
\definecolor{cornflowerblue}{rgb}{0.39,0.58,0.93}
\definecolor{cornsilk1}{rgb}{1.00,0.97,0.86}
\definecolor{cornsilk2}{rgb}{0.93,0.91,0.80}
\definecolor{cornsilk3}{rgb}{0.80,0.78,0.69}
\definecolor{cornsilk4}{rgb}{0.55,0.53,0.47}
\definecolor{cornsilk}{rgb}{1.00,0.97,0.86}
\definecolor{cyan1}{rgb}{0.00,1.00,1.00}
\definecolor{cyan2}{rgb}{0.00,0.93,0.93}
\definecolor{cyan3}{rgb}{0.00,0.80,0.80}
\definecolor{cyan4}{rgb}{0.00,0.55,0.55}
\definecolor{cyan}{rgb}{0.00,1.00,1.00}
\definecolor{darkblue}{rgb}{0.00,0.00,0.55}
\definecolor{darkcyan}{rgb}{0.00,0.55,0.55}
\definecolor{darkgoldenrod}{rgb}{0.72,0.53,0.04}
\definecolor{darkgray}{rgb}{0.66,0.66,0.66}
\definecolor{darkgreen}{rgb}{0.00,0.39,0.00}
\definecolor{darkgrey}{rgb}{0.66,0.66,0.66}
\definecolor{darkkhaki}{rgb}{0.74,0.72,0.42}
\definecolor{darkmagenta}{rgb}{0.55,0.00,0.55}
\definecolor{darkolive}{rgb}{0.33,0.42,0.18}
\definecolor{darkorange}{rgb}{1.00,0.55,0.00}
\definecolor{darkorchid}{rgb}{0.60,0.20,0.80}
\definecolor{darkred}{rgb}{0.55,0.00,0.00}
\definecolor{darksalmon}{rgb}{0.91,0.59,0.48}
\definecolor{darksea}{rgb}{0.56,0.74,0.56}
\definecolor{darkslate}{rgb}{0.18,0.31,0.31}
\definecolor{darkslate}{rgb}{0.18,0.31,0.31}
\definecolor{darkslate}{rgb}{0.28,0.24,0.55}
\definecolor{darkturquoise}{rgb}{0.00,0.81,0.82}
\definecolor{darkviolet}{rgb}{0.58,0.00,0.83}
\definecolor{deeppink}{rgb}{1.00,0.08,0.58}
\definecolor{deepsky}{rgb}{0.00,0.75,1.00}
\definecolor{dimgray}{rgb}{0.41,0.41,0.41}
\definecolor{dimgrey}{rgb}{0.41,0.41,0.41}
\definecolor{dodgerblue}{rgb}{0.12,0.56,1.00}
\definecolor{firebrick1}{rgb}{1.00,0.19,0.19}
\definecolor{firebrick2}{rgb}{0.93,0.17,0.17}
\definecolor{firebrick3}{rgb}{0.80,0.15,0.15}
\definecolor{firebrick4}{rgb}{0.55,0.10,0.10}
\definecolor{firebrick}{rgb}{0.70,0.13,0.13}
\definecolor{floralwhite}{rgb}{1.00,0.98,0.94}
\definecolor{forestgreen}{rgb}{0.13,0.55,0.13}
\definecolor{gainsboro}{rgb}{0.86,0.86,0.86}
\definecolor{ghostwhite}{rgb}{0.97,0.97,1.00}
\definecolor{gold1}{rgb}{1.00,0.84,0.00}
\definecolor{gold2}{rgb}{0.93,0.79,0.00}
\definecolor{gold3}{rgb}{0.80,0.68,0.00}
\definecolor{gold4}{rgb}{0.55,0.46,0.00}
\definecolor{goldenrod1}{rgb}{1.00,0.76,0.15}
\definecolor{goldenrod2}{rgb}{0.93,0.71,0.13}
\definecolor{goldenrod3}{rgb}{0.80,0.61,0.11}
\definecolor{goldenrod4}{rgb}{0.55,0.41,0.08}
\definecolor{goldenrod}{rgb}{0.85,0.65,0.13}
\definecolor{gold}{rgb}{1.00,0.84,0.00}
\definecolor{gray0}{rgb}{0.00,0.00,0.00}
\definecolor{gray100}{rgb}{1.00,1.00,1.00}
\definecolor{gray10}{rgb}{0.10,0.10,0.10}
\definecolor{gray11}{rgb}{0.11,0.11,0.11}
\definecolor{gray12}{rgb}{0.12,0.12,0.12}
\definecolor{gray13}{rgb}{0.13,0.13,0.13}
\definecolor{gray14}{rgb}{0.14,0.14,0.14}
\definecolor{gray15}{rgb}{0.15,0.15,0.15}
\definecolor{gray16}{rgb}{0.16,0.16,0.16}
\definecolor{gray17}{rgb}{0.17,0.17,0.17}
\definecolor{gray18}{rgb}{0.18,0.18,0.18}
\definecolor{gray19}{rgb}{0.19,0.19,0.19}
\definecolor{gray1}{rgb}{0.01,0.01,0.01}
\definecolor{gray20}{rgb}{0.20,0.20,0.20}
\definecolor{gray21}{rgb}{0.21,0.21,0.21}
\definecolor{gray22}{rgb}{0.22,0.22,0.22}
\definecolor{gray23}{rgb}{0.23,0.23,0.23}
\definecolor{gray24}{rgb}{0.24,0.24,0.24}
\definecolor{gray25}{rgb}{0.25,0.25,0.25}
\definecolor{gray26}{rgb}{0.26,0.26,0.26}
\definecolor{gray27}{rgb}{0.27,0.27,0.27}
\definecolor{gray28}{rgb}{0.28,0.28,0.28}
\definecolor{gray29}{rgb}{0.29,0.29,0.29}
\definecolor{gray2}{rgb}{0.02,0.02,0.02}
\definecolor{gray30}{rgb}{0.30,0.30,0.30}
\definecolor{gray31}{rgb}{0.31,0.31,0.31}
\definecolor{gray32}{rgb}{0.32,0.32,0.32}
\definecolor{gray33}{rgb}{0.33,0.33,0.33}
\definecolor{gray34}{rgb}{0.34,0.34,0.34}
\definecolor{gray35}{rgb}{0.35,0.35,0.35}
\definecolor{gray36}{rgb}{0.36,0.36,0.36}
\definecolor{gray37}{rgb}{0.37,0.37,0.37}
\definecolor{gray38}{rgb}{0.38,0.38,0.38}
\definecolor{gray39}{rgb}{0.39,0.39,0.39}
\definecolor{gray3}{rgb}{0.03,0.03,0.03}
\definecolor{gray40}{rgb}{0.40,0.40,0.40}
\definecolor{gray41}{rgb}{0.41,0.41,0.41}
\definecolor{gray42}{rgb}{0.42,0.42,0.42}
\definecolor{gray43}{rgb}{0.43,0.43,0.43}
\definecolor{gray44}{rgb}{0.44,0.44,0.44}
\definecolor{gray45}{rgb}{0.45,0.45,0.45}
\definecolor{gray46}{rgb}{0.46,0.46,0.46}
\definecolor{gray47}{rgb}{0.47,0.47,0.47}
\definecolor{gray48}{rgb}{0.48,0.48,0.48}
\definecolor{gray49}{rgb}{0.49,0.49,0.49}
\definecolor{gray4}{rgb}{0.04,0.04,0.04}
\definecolor{gray50}{rgb}{0.50,0.50,0.50}
\definecolor{gray51}{rgb}{0.51,0.51,0.51}
\definecolor{gray52}{rgb}{0.52,0.52,0.52}
\definecolor{gray53}{rgb}{0.53,0.53,0.53}
\definecolor{gray54}{rgb}{0.54,0.54,0.54}
\definecolor{gray55}{rgb}{0.55,0.55,0.55}
\definecolor{gray56}{rgb}{0.56,0.56,0.56}
\definecolor{gray57}{rgb}{0.57,0.57,0.57}
\definecolor{gray58}{rgb}{0.58,0.58,0.58}
\definecolor{gray59}{rgb}{0.59,0.59,0.59}
\definecolor{gray5}{rgb}{0.05,0.05,0.05}
\definecolor{gray60}{rgb}{0.60,0.60,0.60}
\definecolor{gray61}{rgb}{0.61,0.61,0.61}
\definecolor{gray62}{rgb}{0.62,0.62,0.62}
\definecolor{gray63}{rgb}{0.63,0.63,0.63}
\definecolor{gray64}{rgb}{0.64,0.64,0.64}
\definecolor{gray65}{rgb}{0.65,0.65,0.65}
\definecolor{gray66}{rgb}{0.66,0.66,0.66}
\definecolor{gray67}{rgb}{0.67,0.67,0.67}
\definecolor{gray68}{rgb}{0.68,0.68,0.68}
\definecolor{gray69}{rgb}{0.69,0.69,0.69}
\definecolor{gray6}{rgb}{0.06,0.06,0.06}
\definecolor{gray70}{rgb}{0.70,0.70,0.70}
\definecolor{gray71}{rgb}{0.71,0.71,0.71}
\definecolor{gray72}{rgb}{0.72,0.72,0.72}
\definecolor{gray73}{rgb}{0.73,0.73,0.73}
\definecolor{gray74}{rgb}{0.74,0.74,0.74}
\definecolor{gray75}{rgb}{0.75,0.75,0.75}
\definecolor{gray76}{rgb}{0.76,0.76,0.76}
\definecolor{gray77}{rgb}{0.77,0.77,0.77}
\definecolor{gray78}{rgb}{0.78,0.78,0.78}
\definecolor{gray79}{rgb}{0.79,0.79,0.79}
\definecolor{gray7}{rgb}{0.07,0.07,0.07}
\definecolor{gray80}{rgb}{0.80,0.80,0.80}
\definecolor{gray81}{rgb}{0.81,0.81,0.81}
\definecolor{gray82}{rgb}{0.82,0.82,0.82}
\definecolor{gray83}{rgb}{0.83,0.83,0.83}
\definecolor{gray84}{rgb}{0.84,0.84,0.84}
\definecolor{gray85}{rgb}{0.85,0.85,0.85}
\definecolor{gray86}{rgb}{0.86,0.86,0.86}
\definecolor{gray87}{rgb}{0.87,0.87,0.87}
\definecolor{gray88}{rgb}{0.88,0.88,0.88}
\definecolor{gray89}{rgb}{0.89,0.89,0.89}
\definecolor{gray8}{rgb}{0.08,0.08,0.08}
\definecolor{gray90}{rgb}{0.90,0.90,0.90}
\definecolor{gray91}{rgb}{0.91,0.91,0.91}
\definecolor{gray92}{rgb}{0.92,0.92,0.92}
\definecolor{gray93}{rgb}{0.93,0.93,0.93}
\definecolor{gray94}{rgb}{0.94,0.94,0.94}
\definecolor{gray95}{rgb}{0.95,0.95,0.95}
\definecolor{gray96}{rgb}{0.96,0.96,0.96}
\definecolor{gray97}{rgb}{0.97,0.97,0.97}
\definecolor{gray98}{rgb}{0.98,0.98,0.98}
\definecolor{gray99}{rgb}{0.99,0.99,0.99}
\definecolor{gray9}{rgb}{0.09,0.09,0.09}
\definecolor{gray}{rgb}{0.75,0.75,0.75}
\definecolor{green1}{rgb}{0.00,1.00,0.00}
\definecolor{green2}{rgb}{0.00,0.93,0.00}
\definecolor{green3}{rgb}{0.00,0.80,0.00}
\definecolor{green4}{rgb}{0.00,0.55,0.00}
\definecolor{greenyellow}{rgb}{0.68,1.00,0.18}
\definecolor{green}{rgb}{0.00,1.00,0.00}
\definecolor{grey0}{rgb}{0.00,0.00,0.00}
\definecolor{grey100}{rgb}{1.00,1.00,1.00}
\definecolor{grey10}{rgb}{0.10,0.10,0.10}
\definecolor{grey11}{rgb}{0.11,0.11,0.11}
\definecolor{grey12}{rgb}{0.12,0.12,0.12}
\definecolor{grey13}{rgb}{0.13,0.13,0.13}
\definecolor{grey14}{rgb}{0.14,0.14,0.14}
\definecolor{grey15}{rgb}{0.15,0.15,0.15}
\definecolor{grey16}{rgb}{0.16,0.16,0.16}
\definecolor{grey17}{rgb}{0.17,0.17,0.17}
\definecolor{grey18}{rgb}{0.18,0.18,0.18}
\definecolor{grey19}{rgb}{0.19,0.19,0.19}
\definecolor{grey1}{rgb}{0.01,0.01,0.01}
\definecolor{grey20}{rgb}{0.20,0.20,0.20}
\definecolor{grey21}{rgb}{0.21,0.21,0.21}
\definecolor{grey22}{rgb}{0.22,0.22,0.22}
\definecolor{grey23}{rgb}{0.23,0.23,0.23}
\definecolor{grey24}{rgb}{0.24,0.24,0.24}
\definecolor{grey25}{rgb}{0.25,0.25,0.25}
\definecolor{grey26}{rgb}{0.26,0.26,0.26}
\definecolor{grey27}{rgb}{0.27,0.27,0.27}
\definecolor{grey28}{rgb}{0.28,0.28,0.28}
\definecolor{grey29}{rgb}{0.29,0.29,0.29}
\definecolor{grey2}{rgb}{0.02,0.02,0.02}
\definecolor{grey30}{rgb}{0.30,0.30,0.30}
\definecolor{grey31}{rgb}{0.31,0.31,0.31}
\definecolor{grey32}{rgb}{0.32,0.32,0.32}
\definecolor{grey33}{rgb}{0.33,0.33,0.33}
\definecolor{grey34}{rgb}{0.34,0.34,0.34}
\definecolor{grey35}{rgb}{0.35,0.35,0.35}
\definecolor{grey36}{rgb}{0.36,0.36,0.36}
\definecolor{grey37}{rgb}{0.37,0.37,0.37}
\definecolor{grey38}{rgb}{0.38,0.38,0.38}
\definecolor{grey39}{rgb}{0.39,0.39,0.39}
\definecolor{grey3}{rgb}{0.03,0.03,0.03}
\definecolor{grey40}{rgb}{0.40,0.40,0.40}
\definecolor{grey41}{rgb}{0.41,0.41,0.41}
\definecolor{grey42}{rgb}{0.42,0.42,0.42}
\definecolor{grey43}{rgb}{0.43,0.43,0.43}
\definecolor{grey44}{rgb}{0.44,0.44,0.44}
\definecolor{grey45}{rgb}{0.45,0.45,0.45}
\definecolor{grey46}{rgb}{0.46,0.46,0.46}
\definecolor{grey47}{rgb}{0.47,0.47,0.47}
\definecolor{grey48}{rgb}{0.48,0.48,0.48}
\definecolor{grey49}{rgb}{0.49,0.49,0.49}
\definecolor{grey4}{rgb}{0.04,0.04,0.04}
\definecolor{grey50}{rgb}{0.50,0.50,0.50}
\definecolor{grey51}{rgb}{0.51,0.51,0.51}
\definecolor{grey52}{rgb}{0.52,0.52,0.52}
\definecolor{grey53}{rgb}{0.53,0.53,0.53}
\definecolor{grey54}{rgb}{0.54,0.54,0.54}
\definecolor{grey55}{rgb}{0.55,0.55,0.55}
\definecolor{grey56}{rgb}{0.56,0.56,0.56}
\definecolor{grey57}{rgb}{0.57,0.57,0.57}
\definecolor{grey58}{rgb}{0.58,0.58,0.58}
\definecolor{grey59}{rgb}{0.59,0.59,0.59}
\definecolor{grey5}{rgb}{0.05,0.05,0.05}
\definecolor{grey60}{rgb}{0.60,0.60,0.60}
\definecolor{grey61}{rgb}{0.61,0.61,0.61}
\definecolor{grey62}{rgb}{0.62,0.62,0.62}
\definecolor{grey63}{rgb}{0.63,0.63,0.63}
\definecolor{grey64}{rgb}{0.64,0.64,0.64}
\definecolor{grey65}{rgb}{0.65,0.65,0.65}
\definecolor{grey66}{rgb}{0.66,0.66,0.66}
\definecolor{grey67}{rgb}{0.67,0.67,0.67}
\definecolor{grey68}{rgb}{0.68,0.68,0.68}
\definecolor{grey69}{rgb}{0.69,0.69,0.69}
\definecolor{grey6}{rgb}{0.06,0.06,0.06}
\definecolor{grey70}{rgb}{0.70,0.70,0.70}
\definecolor{grey71}{rgb}{0.71,0.71,0.71}
\definecolor{grey72}{rgb}{0.72,0.72,0.72}
\definecolor{grey73}{rgb}{0.73,0.73,0.73}
\definecolor{grey74}{rgb}{0.74,0.74,0.74}
\definecolor{grey75}{rgb}{0.75,0.75,0.75}
\definecolor{grey76}{rgb}{0.76,0.76,0.76}
\definecolor{grey77}{rgb}{0.77,0.77,0.77}
\definecolor{grey78}{rgb}{0.78,0.78,0.78}
\definecolor{grey79}{rgb}{0.79,0.79,0.79}
\definecolor{grey7}{rgb}{0.07,0.07,0.07}
\definecolor{grey80}{rgb}{0.80,0.80,0.80}
\definecolor{grey81}{rgb}{0.81,0.81,0.81}
\definecolor{grey82}{rgb}{0.82,0.82,0.82}
\definecolor{grey83}{rgb}{0.83,0.83,0.83}
\definecolor{grey84}{rgb}{0.84,0.84,0.84}
\definecolor{grey85}{rgb}{0.85,0.85,0.85}
\definecolor{grey86}{rgb}{0.86,0.86,0.86}
\definecolor{grey87}{rgb}{0.87,0.87,0.87}
\definecolor{grey88}{rgb}{0.88,0.88,0.88}
\definecolor{grey89}{rgb}{0.89,0.89,0.89}
\definecolor{grey8}{rgb}{0.08,0.08,0.08}
\definecolor{grey90}{rgb}{0.90,0.90,0.90}
\definecolor{grey91}{rgb}{0.91,0.91,0.91}
\definecolor{grey92}{rgb}{0.92,0.92,0.92}
\definecolor{grey93}{rgb}{0.93,0.93,0.93}
\definecolor{grey94}{rgb}{0.94,0.94,0.94}
\definecolor{grey95}{rgb}{0.95,0.95,0.95}
\definecolor{grey96}{rgb}{0.96,0.96,0.96}
\definecolor{grey97}{rgb}{0.97,0.97,0.97}
\definecolor{grey98}{rgb}{0.98,0.98,0.98}
\definecolor{grey99}{rgb}{0.99,0.99,0.99}
\definecolor{grey9}{rgb}{0.09,0.09,0.09}
\definecolor{grey}{rgb}{0.75,0.75,0.75}
\definecolor{honeydew1}{rgb}{0.94,1.00,0.94}
\definecolor{honeydew2}{rgb}{0.88,0.93,0.88}
\definecolor{honeydew3}{rgb}{0.76,0.80,0.76}
\definecolor{honeydew4}{rgb}{0.51,0.55,0.51}
\definecolor{honeydew}{rgb}{0.94,1.00,0.94}
\definecolor{hotpink}{rgb}{1.00,0.41,0.71}
\definecolor{indianred}{rgb}{0.80,0.36,0.36}
\definecolor{ivory1}{rgb}{1.00,1.00,0.94}
\definecolor{ivory2}{rgb}{0.93,0.93,0.88}
\definecolor{ivory3}{rgb}{0.80,0.80,0.76}
\definecolor{ivory4}{rgb}{0.55,0.55,0.51}
\definecolor{ivory}{rgb}{1.00,1.00,0.94}
\definecolor{khaki1}{rgb}{1.00,0.96,0.56}
\definecolor{khaki2}{rgb}{0.93,0.90,0.52}
\definecolor{khaki3}{rgb}{0.80,0.78,0.45}
\definecolor{khaki4}{rgb}{0.55,0.53,0.31}
\definecolor{khaki}{rgb}{0.94,0.90,0.55}
\definecolor{lavenderblush}{rgb}{1.00,0.94,0.96}
\definecolor{lavender}{rgb}{0.90,0.90,0.98}
\definecolor{lawngreen}{rgb}{0.49,0.99,0.00}
\definecolor{lemonchiffon}{rgb}{1.00,0.98,0.80}
\definecolor{lightblue}{rgb}{0.68,0.85,0.90}
\definecolor{lightcoral}{rgb}{0.94,0.50,0.50}
\definecolor{lightcyan}{rgb}{0.88,1.00,1.00}
\definecolor{lightgoldenrod}{rgb}{0.93,0.87,0.51}
\definecolor{lightgoldenrod}{rgb}{0.98,0.98,0.82}
\definecolor{lightgray}{rgb}{0.83,0.83,0.83}
\definecolor{lightgreen}{rgb}{0.56,0.93,0.56}
\definecolor{lightgrey}{rgb}{0.83,0.83,0.83}
\definecolor{lightpink}{rgb}{1.00,0.71,0.76}
\definecolor{lightsalmon}{rgb}{1.00,0.63,0.48}
\definecolor{lightsea}{rgb}{0.13,0.70,0.67}
\definecolor{lightsky}{rgb}{0.53,0.81,0.98}
\definecolor{lightslate}{rgb}{0.47,0.53,0.60}
\definecolor{lightslate}{rgb}{0.47,0.53,0.60}
\definecolor{lightslate}{rgb}{0.52,0.44,1.00}
\definecolor{lightsteel}{rgb}{0.69,0.77,0.87}
\definecolor{lightyellow}{rgb}{1.00,1.00,0.88}
\definecolor{limegreen}{rgb}{0.20,0.80,0.20}
\definecolor{linen}{rgb}{0.98,0.94,0.90}
\definecolor{magenta1}{rgb}{1.00,0.00,1.00}
\definecolor{magenta2}{rgb}{0.93,0.00,0.93}
\definecolor{magenta3}{rgb}{0.80,0.00,0.80}
\definecolor{magenta4}{rgb}{0.55,0.00,0.55}
\definecolor{magenta}{rgb}{1.00,0.00,1.00}
\definecolor{maroon1}{rgb}{1.00,0.20,0.70}
\definecolor{maroon2}{rgb}{0.93,0.19,0.65}
\definecolor{maroon3}{rgb}{0.80,0.16,0.56}
\definecolor{maroon4}{rgb}{0.55,0.11,0.38}
\definecolor{maroon}{rgb}{0.69,0.19,0.38}
\definecolor{mediumaquamarine}{rgb}{0.40,0.80,0.67}
\definecolor{mediumblue}{rgb}{0.00,0.00,0.80}
\definecolor{mediumorchid}{rgb}{0.73,0.33,0.83}
\definecolor{mediumpurple}{rgb}{0.58,0.44,0.86}
\definecolor{mediumsea}{rgb}{0.24,0.70,0.44}
\definecolor{mediumslate}{rgb}{0.48,0.41,0.93}
\definecolor{mediumspring}{rgb}{0.00,0.98,0.60}
\definecolor{mediumturquoise}{rgb}{0.28,0.82,0.80}
\definecolor{mediumviolet}{rgb}{0.78,0.08,0.52}
\definecolor{midnightblue}{rgb}{0.10,0.10,0.44}
\definecolor{mintcream}{rgb}{0.96,1.00,0.98}
\definecolor{mistyrose}{rgb}{1.00,0.89,0.88}
\definecolor{moccasin}{rgb}{1.00,0.89,0.71}
\definecolor{navajowhite}{rgb}{1.00,0.87,0.68}
\definecolor{navyblue}{rgb}{0.00,0.00,0.50}
\definecolor{navy}{rgb}{0.00,0.00,0.50}
\definecolor{oldlace}{rgb}{0.99,0.96,0.90}
\definecolor{olivedrab}{rgb}{0.42,0.56,0.14}
\definecolor{orange1}{rgb}{1.00,0.65,0.00}
\definecolor{orange2}{rgb}{0.93,0.60,0.00}
\definecolor{orange3}{rgb}{0.80,0.52,0.00}
\definecolor{orange4}{rgb}{0.55,0.35,0.00}
\definecolor{orangered}{rgb}{1.00,0.27,0.00}
\definecolor{orange}{rgb}{1.00,0.65,0.00}
\definecolor{orchid1}{rgb}{1.00,0.51,0.98}
\definecolor{orchid2}{rgb}{0.93,0.48,0.91}
\definecolor{orchid3}{rgb}{0.80,0.41,0.79}
\definecolor{orchid4}{rgb}{0.55,0.28,0.54}
\definecolor{orchid}{rgb}{0.85,0.44,0.84}
\definecolor{palegoldenrod}{rgb}{0.93,0.91,0.67}
\definecolor{palegreen}{rgb}{0.60,0.98,0.60}
\definecolor{paleturquoise}{rgb}{0.69,0.93,0.93}
\definecolor{paleviolet}{rgb}{0.86,0.44,0.58}
\definecolor{papayawhip}{rgb}{1.00,0.94,0.84}
\definecolor{peachpuff}{rgb}{1.00,0.85,0.73}
\definecolor{peru}{rgb}{0.80,0.52,0.25}
\definecolor{pink1}{rgb}{1.00,0.71,0.77}
\definecolor{pink2}{rgb}{0.93,0.66,0.72}
\definecolor{pink3}{rgb}{0.80,0.57,0.62}
\definecolor{pink4}{rgb}{0.55,0.39,0.42}
\definecolor{pink}{rgb}{1.00,0.75,0.80}
\definecolor{plum1}{rgb}{1.00,0.73,1.00}
\definecolor{plum2}{rgb}{0.93,0.68,0.93}
\definecolor{plum3}{rgb}{0.80,0.59,0.80}
\definecolor{plum4}{rgb}{0.55,0.40,0.55}
\definecolor{plum}{rgb}{0.87,0.63,0.87}
\definecolor{powderblue}{rgb}{0.69,0.88,0.90}
\definecolor{purple1}{rgb}{0.61,0.19,1.00}
\definecolor{purple2}{rgb}{0.57,0.17,0.93}
\definecolor{purple3}{rgb}{0.49,0.15,0.80}
\definecolor{purple4}{rgb}{0.33,0.10,0.55}
\definecolor{purple}{rgb}{0.63,0.13,0.94}
\definecolor{red1}{rgb}{1.00,0.00,0.00}
\definecolor{red2}{rgb}{0.93,0.00,0.00}
\definecolor{red3}{rgb}{0.80,0.00,0.00}
\definecolor{red4}{rgb}{0.55,0.00,0.00}
\definecolor{red}{rgb}{1.00,0.00,0.00}
\definecolor{rosybrown}{rgb}{0.74,0.56,0.56}
\definecolor{royalblue}{rgb}{0.25,0.41,0.88}
\definecolor{saddlebrown}{rgb}{0.55,0.27,0.07}
\definecolor{salmon1}{rgb}{1.00,0.55,0.41}
\definecolor{salmon2}{rgb}{0.93,0.51,0.38}
\definecolor{salmon3}{rgb}{0.80,0.44,0.33}
\definecolor{salmon4}{rgb}{0.55,0.30,0.22}
\definecolor{salmon}{rgb}{0.98,0.50,0.45}
\definecolor{sandybrown}{rgb}{0.96,0.64,0.38}
\definecolor{seagreen}{rgb}{0.18,0.55,0.34}
\definecolor{seashell1}{rgb}{1.00,0.96,0.93}
\definecolor{seashell2}{rgb}{0.93,0.90,0.87}
\definecolor{seashell3}{rgb}{0.80,0.77,0.75}
\definecolor{seashell4}{rgb}{0.55,0.53,0.51}
\definecolor{seashell}{rgb}{1.00,0.96,0.93}
\definecolor{sienna1}{rgb}{1.00,0.51,0.28}
\definecolor{sienna2}{rgb}{0.93,0.47,0.26}
\definecolor{sienna3}{rgb}{0.80,0.41,0.22}
\definecolor{sienna4}{rgb}{0.55,0.28,0.15}
\definecolor{sienna}{rgb}{0.63,0.32,0.18}
\definecolor{skyblue}{rgb}{0.53,0.81,0.92}
\definecolor{slateblue}{rgb}{0.42,0.35,0.80}
\definecolor{slategray}{rgb}{0.44,0.50,0.56}
\definecolor{slategrey}{rgb}{0.44,0.50,0.56}
\definecolor{snow1}{rgb}{1.00,0.98,0.98}
\definecolor{snow2}{rgb}{0.93,0.91,0.91}
\definecolor{snow3}{rgb}{0.80,0.79,0.79}
\definecolor{snow4}{rgb}{0.55,0.54,0.54}
\definecolor{snow}{rgb}{1.00,0.98,0.98}
\definecolor{springgreen}{rgb}{0.00,1.00,0.50}
\definecolor{steelblue}{rgb}{0.27,0.51,0.71}
\definecolor{tan1}{rgb}{1.00,0.65,0.31}
\definecolor{tan2}{rgb}{0.93,0.60,0.29}
\definecolor{tan3}{rgb}{0.80,0.52,0.25}
\definecolor{tan4}{rgb}{0.55,0.35,0.17}
\definecolor{tan}{rgb}{0.82,0.71,0.55}
\definecolor{thistle1}{rgb}{1.00,0.88,1.00}
\definecolor{thistle2}{rgb}{0.93,0.82,0.93}
\definecolor{thistle3}{rgb}{0.80,0.71,0.80}
\definecolor{thistle4}{rgb}{0.55,0.48,0.55}
\definecolor{thistle}{rgb}{0.85,0.75,0.85}
\definecolor{tomato1}{rgb}{1.00,0.39,0.28}
\definecolor{tomato2}{rgb}{0.93,0.36,0.26}
\definecolor{tomato3}{rgb}{0.80,0.31,0.22}
\definecolor{tomato4}{rgb}{0.55,0.21,0.15}
\definecolor{tomato}{rgb}{1.00,0.39,0.28}
\definecolor{turquoise1}{rgb}{0.00,0.96,1.00}
\definecolor{turquoise2}{rgb}{0.00,0.90,0.93}
\definecolor{turquoise3}{rgb}{0.00,0.77,0.80}
\definecolor{turquoise4}{rgb}{0.00,0.53,0.55}
\definecolor{turquoise}{rgb}{0.25,0.88,0.82}
\definecolor{violetred}{rgb}{0.82,0.13,0.56}
\definecolor{violet}{rgb}{0.93,0.51,0.93}
\definecolor{wheat1}{rgb}{1.00,0.91,0.73}
\definecolor{wheat2}{rgb}{0.93,0.85,0.68}
\definecolor{wheat3}{rgb}{0.80,0.73,0.59}
\definecolor{wheat4}{rgb}{0.55,0.49,0.40}
\definecolor{wheat}{rgb}{0.96,0.87,0.70}
\definecolor{whitesmoke}{rgb}{0.96,0.96,0.96}
\definecolor{white}{rgb}{1.00,1.00,1.00}
\definecolor{yellow1}{rgb}{1.00,1.00,0.00}
\definecolor{yellow2}{rgb}{0.93,0.93,0.00}
\definecolor{yellow3}{rgb}{0.80,0.80,0.00}
\definecolor{yellow4}{rgb}{0.55,0.55,0.00}
\definecolor{yellowgreen}{rgb}{0.60,0.80,0.20}
\definecolor{yellow}{rgb}{1.00,1.00,0.00}

\title[AGN feedback in nearby early-type galaxies]
    {A simple model for AGN feedback in nearby early-type galaxies}

\author[Sugata Kaviraj et al.]
{Sugata Kaviraj$^{1,2}$\thanks{E-mail: s.kaviraj@imperial.ac.uk}\thanks{1851 Fellow and Imperial College Research Fellow}, Kevin Schawinski$^{3}$\thanks{Einstein Fellow}, Joseph Silk$^{2}$ and Stanislav S. Shabala$^{2}$\\
$^{1}$Blackett Laboratory, Imperial College London, London SW7 2AZ, UK\\
$^{2}$Department of Physics, University of Oxford, Keble Road, Oxford, OX1 3RH, UK\\
$^{3}$Yale Center for Astronomy and Astrophysics, Yale University,
P.O. Box 208121, New Haven, CT 06520, U.S.A.}

\begin{document}

\vspace{-1.5in}

\date{6 May 2010}
\maketitle

\label{firstpage}


\begin{abstract}
Recent work indicates that star-forming early-type galaxies
residing in the blue cloud migrate rapidly to the red sequence
within around a Gyr, passing through several phases of
increasingly strong AGN activity in the process (Schawinski et al.
2007, MNRAS, 382, 1415; S07 hereafter). We show that natural
depletion of the {\color{black}cold} gas reservoir through star
formation (i.e. in the absence of any feedback from the AGN)
induces a blue-to-red reddening rate that is several factors lower
than that observed in S07. This is because the gas depletion rate
due to star formation alone is too slow, implying that another
process needs to be invoked to remove {\color{black}cold} gas from
the system and accelerate the reddening rate. We develop a simple
phenomenological model, in which a fraction of the AGN's
luminosity couples to the gas reservoir over a certain `feedback
timescale' and removes part of the {\color{black}cold} gas mass
from the galaxy, while the remaining gas continues to contribute
to star formation. We use the model to investigate scenarios which
yield migration times consistent with the results of S07. We find
that acceptable models have feedback timescales $\lesssim 0.2$
Gyrs. The mass fraction in young stars in the remnants is
$\lesssim 5$\% and the residual {\color{black}cold} gas fractions
are less than 0.6\%, in good agreement with the recent literature.
At least half of the initial {\color{black}cold} gas reservoir is
removed as the galaxies evolve from the blue cloud to the red
sequence. If we restrict ourselves to feedback timescales similar
to the typical duty cycles of local AGN (a few hundred Myrs) then
a few tenths of a percent of the luminosity of an early-type
Seyfert ($\sim 10^{11}$ L$_{\odot}$) must couple to the
{\color{black}cold} gas reservoir in order to produce migration
times that are consistent with the observations.
\end{abstract}




\section{Introduction}
The development of the current generation of galaxy formation
models has been inextricably linked to our understanding of the
properties of early-type galaxies. Their red optical colours
\citep[e.g.][]{BLE92,Ellis1997,VD2000,Bernardi2003a,Bell2004,Faber2007},
high alpha-enhancement ratios
\citep[e.g.][]{Thomas1999,Trager2000a,Trager2000b,Thomas2005} and
their obedience of a tight `Fundamental Plane'
\citep[e.g.][]{Jorg1996,Saglia1997,Forbes1998,Peebles2002,Franx1993,Franx1995,VD1996}
indicate that the bulk ($>$80\%) of their constituent stellar mass
forms at high redshift ($z>1$). The star formation at late epochs
\citep[e.g.][]{Trager2000a,Nelan2005,Graves2009a,Graves2009b,Scott2009,VD2010},
recently quantified using rest-frame UV/optical photometry
\citep[][]{Ferreras2000,Yi2005,Kaviraj2007b,Schawinski2007a,Jeong2007,Kaviraj2008a,Kaviraj2008b,Jeong2009,Salim2010},
is plausibly driven by minor merging through the accretion of
gas-rich satellites
(\citealt{Schweizer1990,Schweizer1992,Kaviraj2009,Kaviraj2010a,Kaviraj2010b},
see also
\citealt{Tal2009,Bournaud2007,Bezanson2009,Naab2009,Serra2010,Hopkins2010,Schawinski2010}).
This is supported by evidence for kinematical decoupling of the
(ionised) gas from the stars \citep[e.g.][]{Sarzi2006} and the
fact that the gas and associated dust appears not to correlate
with the stellar mass of the galaxy, irrespective of the local
environment \citep[e.g.][]{VD1995,Knapp1996,Combes2007}. Both
these trends indicate that the gas is, at least in part, external
in origin.

While the characteristics of the {\color{black}early-type galaxy}
population have been exhaustively studied, the reproduction of
those properties in the models remains problematic. A particular
issue has been the continuing availability of cold gas and
resultant star formation in massive galaxies (which are dominated
by {\color{black}early-types}) at late epochs, as supernova
feedback becomes ineffective in very deep potential wells
\citep[see e.g.][]{Dekel1986,Benson2003}. Strong gas cooling on to
the central galaxies of groups and clusters leads to model
galaxies being both too massive and too blue
\citep[e.g.][]{Kauffmann1999,Somerville1999,Cole2000,Murali2002,Benson2003},
with alpha-enhancements that are too low to match the observed
values \citep[e.g.][but see Pipino et al.
2009]{Thomas1999,Nagashima2005}. An additional source of energy is
thus required to prevent cold gas from forming stars, either by
supplementing the heating of the {\color{black}cold} gas reservoir
or, more plausibly, by removing a significant fraction of the
{\color{black}cold} gas mass from the potential well
\citep[e.g][]{Martin1999,Strickland2000}.

Given the ubiquity of super-massive black holes (SMBHs) in local
galaxies \citep[e.g.][]{Richstone1998} and the strong observed
correlation between the masses of SMBHs and the
luminosities/velocity dispersions of their host galaxy bulges
\citep[e.g.][]{Magorrian1998,Gebhardt2000,Ferrarese2000,Tremaine2002,Haring2004},
it is likely that the evolution of galaxies is intimately linked
to, or even regulated by, their central black holes
\citep[e.g][]{Kauffmann2009,Netzer2009}. Consensus now favours
Active Galactic Nuclei (AGN), powered by accretion of matter on to
these central SMBHs, as a potential source of the `missing' energy
that is required to fulfil the feedback budget in massive galaxies
(\citealt{Silk1998,Blandford1999,Fabian1999,Binney2004,Silk2005},
but see also alternatives for imparting energy to the ambient gas
in \citealt{Birnboim2007,Khochfar2008}).

Several processes - e.g. radiative heating or kinetic energy input
through jets \citep[see the recent reviews
by][]{Begelman2004,Fabian2010} may contribute to the deposition of
energy from the AGN into its ambient medium. In powerful AGN, jets
inflate cocoons of relativistic plasma which are overpressured
with respect to the surrounding gas, driving massive outflows
\citep[e.g.][]{Begelman1984,Fabian2006}. While an objection to
invoking this mode of feedback to remove gas from the galaxy is
the small volume-filling factor of the jets
\citep[e.g.][]{Ostriker2005}, recent numerical simulations
indicate a significant (and largely isotropic) interaction between
the jet material and the multi-phase ISM
\citep{Antonuccio_Delogu2008,Sutherland2007}. In any case,
radiative pressure resulting from heating produces similar
momentum-driven outflows \cite[e.g.][]{Ostriker2010} to those
expected in jet-dominated systems. Theoretical arguments indicate
that outflows are necessary in order to produce the observed
correlations between SMBHs and their host galaxies
\citep[e.g.][]{Ostriker2010}. Coupled with observational evidence
for the commonality of outflows in local AGN
\citep[e.g.][]{deKool1997,Crenshaw1999,Crenshaw2003,Everett2007,Proga2007}
this suggests that a major facet of the feedback process is the
injection of kinetic energy and removal of gas from the
interstellar medium (ISM).

Although significant advances have been made in modelling the
complex interaction of the AGN with the inter-stellar medium
\citep[e.g.][]{Falle1991,Kaiser1997,Kino2005,Alexander2006,Krause2007,Antonuccio_Delogu2008},
the inclusion of AGN feedback in galaxy formation models remains
largely phenomenological. Nevertheless, simple recipes for AGN
feedback
\citep[e.g.][]{Hatton2003,Granato2004,Kaviraj2005a,Bower2006,Springel2005a,Croton2006,deLucia2006,Schawinski2006,Khochfar2006,Cattaneo2006,Cattaneo2007,DiMatteo2007,deLucia2007,Somerville2008,Rettura2010},
have proved a valuable addition to the models, enabling good
reproduction of local galaxy properties such as luminosity
functions, the morphological mix of the Universe and the stellar
populations of early-type galaxies. The trigger and intensity of
feedback is postulated to vary with look-back time, with violent
feedback from a `quasar mode' truncating merger-driven star
formation in the gas-rich Universe at high redshift
\citep[e.g.][]{Springel2005d,DiMatteo2005}, while a more quiescent
`maintenance mode' that probably operates in the gas-poor Universe
at late epochs
\citep[e.g.][]{Best2005,Best2006,Schawinski2006,Schawinski2007b,Khalatyan2008,Kormendy2009}.

{\color{black}While energetic arguments make a compelling
theoretical case for the need for AGN feedback, observational
constraints on this feedback process \emph{in the
{\color{black}early-type population} at late epochs} remain
relatively limited but highly desirable}. In a recent work,
\citet[][S07 hereafter]{Schawinski2007b} studied the potential
impact of AGN on {\color{black}early-type} evolution by exploring
the {\color{black}recent star formation} histories of $\sim$16,000
nearby ($0.05<z<0.1$) {\color{black}early-type galaxies} in the
field, as a function of the type of AGN activity present in these
systems. The galaxies, drawn from the Sloan Digital Sky Survey
\citep[SDSS;][]{SDSS_DR6}, were selected through direct visual
inspection of their SDSS images, which yields a more accurate
morphological selection
\citep[e.g][]{Kaviraj2007b,Fukugita2007,Lintott2008} than methods
based on colours or galaxy spectra. AGN diagnostics were performed
using optical emission line ratios
\citep{Baldwin1981,Veilleux1987,Kauffmann2003,Miller2003,Kewley2006},
separating the {\color{black}early-type} population into galaxies
that were `star-forming', `composites' (which have signatures of
both AGN and star formation), `Seyferts', `LINERs' and `quiescent'
systems. The {\color{black}recent star formation} history in each
galaxy was quantified by fitting Lick absorption indices and
multi-wavelength photometry in the ultraviolet (UV), optical and
near-infrared (NIR) wavelengths (from the GALEX
\citep{Martin2005}, SDSS and 2MASS \citep{Skrutskie2006} surveys
respectively) to a large library of model star formation
histories. The model library was constructed using two bursts of
star formation, the first fixed at high redshift (since the bulk
of the mass in {\color{black}early-types} is known to form at
large look-back times), with the second allowed to vary in age and
mass fraction. Realistic values of dust and metallicity were
employed and the age and mass fraction of the second burst (which
characterises the {\color{black}recent star formation} episode)
were calculated for each galaxy by fitting to the
spectro-photometric data. S07 used their estimates of the
{\color{black}recent star formation} as a `clock' to follow the
migration of {\color{black}early-type galaxies} from the red
sequence to the blue cloud and explore the AGN classes that
galaxies passed through in the course of that migration.

The S07 results strongly suggest that star-forming
{\color{black}early-types} residing in the blue cloud migrate
rapidly to the red sequence within $\sim$ a Gyr, passing through
several phases of increasingly strong AGN activity in the process.
The AGN activity reaches its peak around 0.5 Gyrs \citep[see
also][]{Wild2010}. The `reddening sequence' begins with the
star-forming {\color{black}early-types} which are, on average, the
bluest population, followed by the composites, Seyferts, LINERs
and quiescents in that order \citep[see also][who found similar
results]{Salim2007}. The mass fraction in young stars remains
similar along this sequence, while the age of the
{\color{black}recent star formation} progressively increases.
Furthermore, mm-wavelength observations, from the IRAM 30m
telescope, of a subset of the S07 {\color{black}early-type galaxy}
sample indicates that the {\color{black}cold} molecular gas mass
drops precipitously by an order of magnitude between the
star-forming and LINER phases \citep[][S09
hereafter]{Schawinski2009a}. Given the \emph{coincidence} of
rising AGN activity, the rapid observed evolution in colours and
simultaneous fast removal of the molecular gas mass, it is
reasonable to suggest that the AGN may play a significant role in
the migration of {\color{black}early-types} from the blue cloud to
the red sequence. {\color{black}At this point, it is useful to
note the characteristics of the feedback that might operate in
these nearby early-type galaxies. Recent studies indicate that the
trigger for the weak star formation observed in nearby early-type
systems are gas-rich minor mergers \citep[see
e.g.][]{Kaviraj2010a,Kaviraj2010b}. The feedback envisaged here is
in the form of a jet-driven outflow which acts on the gas in the
galaxy and may quench the star formation. Since the supply of gas
in such minor mergers is relatively small, this outflow-driven
feedback is likely to be much weaker than the violent
`quasar-mode' feedback \citep[e.g.][]{Springel2005a} that operates
in gas-rich major mergers, that plausibly drive the quasar
population at high redshift.}

A robust conclusion about the role of the putative AGN feedback in
the S07 {\color{black}early-types} can only be achieved by
comparing the observed colour transition to what might be expected
in the absence of feedback on the system. If feedback was then
found to be necessary, then the optical and mm-wavelength data
presented in S07 and S09 offer an ideal dataset with which the
broad characteristics of the coupling between the AGN and the
{\color{black}cold} gas reservoir in the host
{\color{black}early-type galaxy} can be characterised and
properties of the feedback constrained. {\color{black}Note that,
throughout this paper, we always refer to the gas mass contained
in the molecular i.e. cold phase, hosted in a disk.}

We begin, in Section 2, by demonstrating that, in the absence of
AGN feedback, the expected blue-to-red colour transition and
associated gas depletion in the star-forming
{\color{black}early-types} is likely to be much slower than that
observed by S07. Proceeding under the assumption that these
processes are accelerated by AGN feedback, we then construct a
simple model, in Section 3, that describes the coupling between
the AGN and the host galaxy's {\color{black}cold} gas reservoir.
In Section 4 we apply this model to a typical star-forming
{\color{black}early-type} in the blue cloud and explore scenarios
which simultaneously reproduce the {\color{black}recent star
formation} observed in local {\color{black}early-type galaxies},
the migration times observed by S07 and the gas depletion history
presented in S09. We use these scenarios to draw general
conclusions about the broad characteristics of the feedback from
the central AGN, in particular the fraction of AGN energy that
must couple to the {\color{black}cold} gas reservoir and the
timescale over which it does so. The novelty of this analysis is
that the model is strongly constrained by these observational
results. Finally, in Section 5, we summarise our findings and
connect our results to recent observational work on
{\color{black}recent star formation} in {\color{black}early-type
galaxies}. The overall aim of this paper is to add an
understanding of the role of AGN feedback to the developing
picture of {\color{black}recent star formation} in
{\color{black}early-types} at late epochs, and provide
observationally-driven constraints on the `maintenance mode' of
AGN feedback that is likely to operate in massive galaxies at low
redshifts.


\section{Evolution in the absence of AGN feedback}
We begin by considering whether natural evolution of the
{\color{black}cold} gas reservoir in star-forming
{\color{black}early-types} can produce the $(u-r)$ colour
transition observed in S07, without the need for invoking
feedback. Star formation {\color{black}depletes} this gas
reservoir which causes the galaxy to redden (assuming it is not
replenished by accretion of fresh gas). To describe this secular
evolution, we appeal to the {\color{black}Schmidt-Kennicutt} law
\citep[e.g][]{Schmidt1959,Kennicutt1998a,Boissier2003,Gao2004},
which describes an apparently universal relationship between star
formation rate (SFR) and {\color{black}cold} gas mass across
almost five decades of gas densities and SFRs in the galaxy
population. Extensively established for disk galaxies and
starbursts \citep[see e.g.][and references
therein]{Kennicutt1998a,Kennicutt1998b}, recent work indicates
that the {\color{black}Schmidt-Kennicutt} law also holds for
{\color{black}early-type galaxies}. In a study of CO emission in
43 representative early-type galaxies from the SAURON survey,
\citet{Combes2007} and \citet{Crocker2010} have shown that
{\color{black}early-types} form a low-SFR extension to the
empirical law in spirals. Given its universality and applicability
to {\color{black}early-type galaxies}, it is reasonable to model
the colour evolution of the star-forming
{\color{black}early-types}, in the absence of feedback, using a
{\color{black}Schmidt-Kennicutt} law.

The {\color{black}Schmidt-Kennicutt} law can be parametrised in
terms of either the gas density or the gas mass. Since we only
have measurements of the {\color{black}cold} gas mass in
{\color{black}early-type galaxies} from S09, it is more relevant
to cast the {\color{black}Schmidt-Kennicutt} law in terms of this
quantity. It is worth noting that star formation recipes in
cosmological models of galaxy formation (e.g. semi-analytical
models), where the galaxies are spatially unresolved, also
commonly employ an {\color{black}Schmidt-Kennicutt} law
parametrised in terms of the gas mass \citep[see
e.g.][]{Somerville1999,Cole2000,Hatton2003,Kaviraj2005a,Bower2006,deLucia2006}.
These star formation recipes, tuned to reproduce the empirical
constraints of \citet{Kennicutt1998a}, enable good reproduction of
the properties of the galaxy population in the local Universe e.g.
the observed luminosity functions in optical filters, the colours
of the local galaxy population and the morphological mix of the
Universe at present day. Following the typical parametrisations
used in models \cite[see e.g.][]{Guiderdoni1998,Hatton2003}, the
{\color{black}Schmidt-Kennicutt} law can be expressed as:

\begin{equation}
\psi = (\epsilon/\tau_{dyn}).M_g,
\end{equation}

where $\psi$ is the star formation rate, $\epsilon$ is the star
formation efficiency, $\tau_{dyn}$ is the dynamical timescale of
the system and $M_g$ is the mass of the {\color{black}cold} gas
reservoir. The observed values of these parameters in the
empirically-determined {\color{black}Schmidt-Kennicutt} law for
spiral disks \citep[][]{Kennicutt1998a,Guiderdoni1998}, indicate
that $\epsilon\sim0.02$ and $\tau_{dyn}\sim0.1$ Gyrs (given
typical dynamical timescales of the gas disks). Before we can
study the expected evolution of {\color{black}early-types} via the
{\color{black}Schmidt-Kennicutt} law, we need to establish typical
values of $\epsilon$, $\tau_{dyn}$ and $M_g$ that are relevant to
the S07 {\color{black}early-type} population.

We assume that the star formation efficiency ($\epsilon$) is the
fiducial 2\% observed in the empirical
{\color{black}Schmidt-Kennicutt} law. Several studies over the
last few decades have shown that giant molecular clouds convert
around 1-2\% of their mass over a dynamical timescale. This result
appears independent of the choice of model for molecular-cloud
lifetimes or evolution and holds irrespective of environment
\citep[see e.g.][and references therein]{Tan2006,Krumholz2007}.
Coupled with the fact that the empirical
{\color{black}Schmidt-Kennicutt} law appears to hold for
{\color{black}early-type galaxies} \citep{Combes2007,Crocker2010},
it seems reasonable to assume the efficiency that underpins this
star formation law. The median dynamical timescale of the S07
{\color{black}early-types}, calculated using their (photometric)
stellar masses and Petrosian radii, is $\sim0.08$ Gyrs. This value
falls within the range of the measured dynamical timescales
(0.05-0.2 Gyrs) of {\color{black}cold}, molecular gas disks
observed in \emph{very} nearby {\color{black}early-type galaxies}
\citep{Young2002}, which presumably drive the {\color{black}recent
star formation}. {\color{black}Note that these dynamical
timescales correspond to galaxies in the local Universe - the
corresponding timescales in the higher redshift Universe are
likely to be smaller}. The cold gas fractions in the star-forming
{\color{black}early-types} can be estimated using the measured
cold gas and stellar masses in S09. The stacked data in S09
indicate that the cold gas fractions remaining in the star-forming
{\color{black}early-types} when they are observed are in the range
5-10\%. Since the mass fractions in young stars that have already
formed in these systems is also a few percent, the initial
{\color{black}cold} gas fractions are likely to be in the range
10-15\%. This is consistent with (and slightly lower than) the gas
fractions that may be inferred from \citet{Kannappan2004}, who
measured the (atomic) gas to stellar mass ratios for SDSS galaxies
using the Hyperleda HI catalogue. Their results for galaxies at
$(u-r)\sim 1.5$ and masses between $10^{10}$ and $10^{11}$
M$_{\odot}$ suggest molecular gas fractions around 15-25\% (after
converting from atomic to molecular gas mass using the
calibrations for {\color{black}early-type galaxies} given by
\citealt{Fukugita1998}). Note, however, that \citet{Kannappan2004}
did not split their galaxies by morphology and that their sample
is certainly skewed towards gas-rich galaxies. Nevertheless, their
results provide a useful sanity check of our assumptions for the
initial {\color{black}cold} gas fractions in our star-forming
{\color{black}early-type galaxy} sample.

We proceed by modelling the evolution of a typical star-forming
{\color{black}early-type} by considering a recent starburst,
evolving according to Eqn. 1, superimposed on an old underlying
stellar population. The underlying population is modelled using a
simple stellar population (SSP) with solar metallicity and an age
of 9 Gyrs. The motivation for an old, solar-metallicity SSP is the
extensive literature on {\color{black}early-type galaxies} which
convincingly demonstrates that the bulk of the stars ($\geq
85-90$\%) in these galaxies form at high redshift, possibly over
short timescales (Bower et al. 1992; Thomas et al. 2005, Kaviraj
et al. 2008a,b) and that the stellar populations in present-day
{\color{black}early-types} are typically metal-rich
\citep{Trager2000a} with a mean value around solar metallicity.
Note that our results are insensitive to small changes in the age
or metallicity of this old SSP, because the old population does
not contribute significantly to the $u$-band flux in a
star-forming {\color{black}early-type galaxy}.

\begin{figure}
\begin{center}
\includegraphics[width=0.5\textwidth]{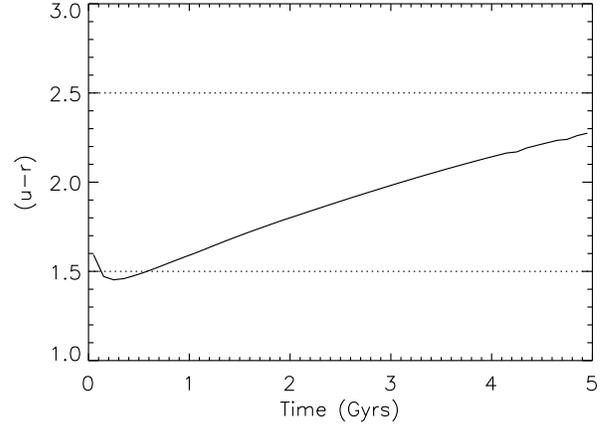}
\caption{The expected evolution of a galaxy according to the
{\color{black}Schmidt-Kennicutt} law (Eqn. 1) where the star
formation efficiency ($\epsilon$) is 0.02, $\tau_{dyn}$ is 0.05
Gyrs and the initial {\color{black}cold} gas fraction is 10\%. The
resultant evolution in the $(u-r)$ colour (in the absence of any
AGN feedback) is too slow to reproduce the rapid transition from
blue cloud to red sequence of {\color{black}early-type galaxies}
in S07. It is worth noting that, given its empirical nature, the
{\color{black}Schmidt-Kennicutt} law includes the impact of
supernovae feedback on the star formation sites in the galaxy.
Note also that the evolution shown here does not take into account
any fresh injection of gas into the system through stellar mass
loss or accretion from the halo. Both of these processes would
further increase the migration time from blue cloud to red
sequence.} \label{fig:sk_evolution}
\end{center}
\end{figure}

In Figure \ref{fig:sk_evolution} we show the expected $(u-r)$
colour evolution of a model {\color{black}early-type} that evolves
according to the {\color{black}Schmidt-Kennicutt} law. We study
the colour evolution between $(u-r)=1.5$ (which represents the
mean colour of the bluest 25\% of the star-forming
{\color{black}early-type} population) and $(u-r)=2.5$ (which
represents the bottom of the red sequence defined by the quiescent
{\color{black}early-type galaxies}). Following the arguments
above, we assume that $\epsilon$ is 0.02, $\tau_{dyn}$ is 0.05
Gyrs and the {\color{black}cold} gas fraction is 10\%. Note that,
to be conservative in our approach, we have chosen the lower limit
for the likely dynamical timescales and {\color{black}cold} gas
fractions in the S07 {\color{black}early-types}, which corresponds
to the fastest possible colour evolution. In principle, if the
colour evolution due to natural evolution of the
{\color{black}cold} gas reservoir is fast enough to be consistent
with the results of S07, then there would be no need for
additional feedback on the system.

Figure \ref{fig:sk_evolution} indicates that the reddening rate
due to pure {\color{black}Schmidt-Kennicutt} evolution [$d(u-r)/dt
\sim 0.16$ Gyr$^{-1}$], is not sufficient to move the galaxy from
the blue cloud ($u-r\sim1.5$) to the bottom of the red sequence
($u-r\sim2.5$) in $\sim 1$ Gyr, as suggested by S07. The reddening
rate is a few factors too slow. It is worth noting that our model
does not assume either stellar mass loss or accretion of gas from
the halo, both of which would slow the depletion of the gas
reservoir and the reddening rate even further. This suggests that,
if the star-forming {\color{black}early-types} were simply
depleting their {\color{black}cold} gas reservoirs through star
formation alone, then they might be rather long-lived blue-cloud
objects, similar to their spiral counterparts at similar colours.
This is not unexpected if the {\color{black}early-types} follow
the same star formation laws as their late-type counterparts as
has been suggested by the studies of \citet{Combes2007} and
\citet{Crocker2010}. Since the observed colour transition is much
faster, it is {\color{black}then} reasonable to suggest that an
additional mechanism needs to be invoked to accelerate the
depletion of available cold gas in the star-forming
{\color{black}early-type galaxies}. {\color{black} It is worth
noting that modern galaxy formation models already incorporate the
\emph{result} of this effect, since massive galaxies in these
models remain too blue in the absence of AGN feedback. The
analysis above isolates this issue and demonstrates explicitly how
the blue-to-red transit times are too long if the gas reservoir is
depleted due to star formation alone.}

The \emph{coincidence} of AGN activity and the rapid observed
evolution in the $(u-r)$ colour strongly suggests that the AGN may
play a significant role in the migration of
{\color{black}early-types} from the blue cloud to the red
sequence. In the following section we develop a simple model, in
which feedback from the AGN accelerates the depletion of the cold
gas reservoir, inducing a faster colour transition that is
consistent with that observed in S07. The characteristics of the
feedback episode are strongly constrained, \emph{observationally},
by the observed migration times in S07, the estimated
{\color{black}mass fractions of young stars} in local
{\color{black}early-type galaxies} from the literature and
residual {\color{black}cold} gas fractions in S09. This allows us
to put some useful constraints on (a) the strength of the coupling
between the AGN and the {\color{black}cold} gas reservoir and (b)
the timescale over which that coupling holds.


\section{A simple model for AGN feedback}
We develop a simple phenomenological model, in which some of the
bolometric luminosity of an AGN couples to the {\color{black}cold
disk gas reservoir and removes some of this gas mass}, thus
accelerating gas depletion and increasing the $(u-r)$ reddening
rate. Given recent observational and theoretical evidence that
outflows may play a dominant role in the AGN feedback process (see
Section 1), the model assumes that {\color{black}cold} gas is
removed from the potential well, motivated by evidence for
momentum-driven outflows contributing significantly to AGN
feedback (see arguments above in the introduction). {\color{black}
As noted in the introduction above, the trigger for the AGN in
nearby early-types is likely to be the accretion of gas-rich
satellites which induces a (weak) jet-driven outflow.}

The coupling between AGN energy and the {\color{black}cold} gas
reservoir is determined by a `feedback function' ($f_t$), which
describes the fraction of the AGN's observed bolometric luminosity
that is deposited into the {\color{black}cold} gas reservoir and
removes a portion of the gas mass. Thus we have, at time $t$:

\begin{equation}
f_t.L_{B}.\delta t=G.M.\delta M_g/R,
\end{equation}

where $L_{B}$ is the observed bolometric luminosity of the AGN,
$G$ is the gravitational constant, $M$ is the mass of the galaxy,
$\delta M_g$ is the {\color{black}cold} gas mass removed, $R$ is
the radius of the galaxy and $\delta t$ is the size of the
timestep being considered.

Note that the left-hand side (LHS) of Eqn. 1 could have been
written simply in terms of the energy deposited into the
{\color{black}cold} gas reservoir i.e. without any reference to
the luminosity of an AGN. In other words, the LHS could be
expressed simply as a luminosity $L_t$, where $L_t \equiv
f_t.L_{B}$. However, our chosen parametrisation allows to us to
cast the feedback energy in terms of a reference luminosity which,
in this case, is the observed luminosity of an AGN. The particular
choice of reference luminosity $L_B$ does not affect the total
feedback energy entering the system of course, it simply allows us
to express the feedback energy in terms of a useful observed
quantity.

\begin{figure}
\begin{center}
\includegraphics[width=0.3\textwidth]{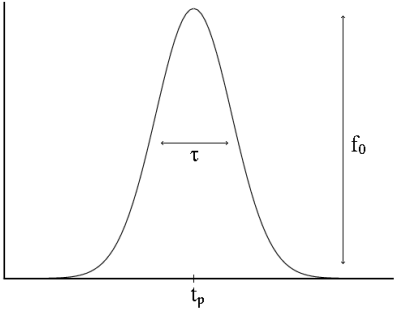}
\caption{A schematic of the feedback function $f_t$. $f_0$ is the
peak fraction of AGN luminosity that couples to the
{\color{black}cold} gas reservoir, $\tau$ is a measure of the
timescale over which the AGN interacts with the
{\color{black}cold} gas reservoir and $t_p$ is the time at which
the coupling is strongest. The results of S07 indicate that $t_p$
is $\sim0.5$ Gyrs.} \label{fig:feedback_schematic}
\end{center}
\end{figure}

The form of $f_t$ (see Figure \ref{fig:feedback_schematic} for a
schematic representation) is assumed to be gaussian,

\begin{equation}
f_t = f_0.\exp\Big[\frac{-(t-t_p)^2}{2\tau^2}\Big],
\end{equation}

with the following free parameters:

\begin{itemize}

    \item $f_0$ is the peak fraction of the luminosity of an AGN that is deposited in the gas
reservoir. It is a measure of how efficiently the AGN couples to
the {\color{black}cold} gas mass in the galaxy, since low values
of $f_0$ will result in less gas being removed from the system and
vice-versa.

    \item $\tau$, the width of the gaussian, is a measure of the timescale over which the AGN interacts
with the {\color{black}cold} gas reservoir. This could, in
principle, involve several AGN episodes over multiple duty cycles.

    \item $t_p$ is the time at which the coupling is strongest. Since the
conclusions in S07 indicate that the AGN activity peaks at roughly
0.5 Gyrs after the onset of star formation, we use a fiducial
value of $t_p\sim0.5$. Small changes to the value of $t_p$ do not
alter our conclusions.

\end{itemize}

With $t_p$ held constant, our primary focus is on exploring the
part of the ($f_0$,$\tau$) parameter space that may reproduce the
reddening rate observed in S07. As noted before, the models are
constrained by three sets of observations - the {\color{black}
recent star formation} produced in the galaxy, the migration time
between the blue cloud and the red sequence and the residual
{\color{black}cold} gas fractions in the galaxies when they arrive
on the red sequence. {\color{black} Note that specifying the shape
of the feedback function as a gaussian is somewhat arbitrary.
However, the S07 results indicate that the AGN activity rises and
falls within a Gyr, with a peak around 0.4 Gyrs. Given the
simplicity of the model, a gaussian function appears a reasonable
way to parametrise the feedback process.}

{\color{black}We note that this is a simple model, decoupled from
cosmological evolution, and geared towards studying star formation
episodes in nearby early-type galaxies, triggered by discrete
minor merger events. While the model is only designed to capture
the broad characteristics of the feedback process, its simplicity
does not allow us to put constraints on details of the feedback
e.g. the number of individual episodes of AGN activity or any
modulation in the AGN output within an episode. Also recall that,
while the model invokes the removal of cold gas - motivated by
evidence for momentum-driven outflows contributing significantly
to AGN feedback - it does not include the effects of radiative
heating which may stop fresh gas cooling onto the disk. It is
worth noting, however, that the early-type galaxies studied here
reside in the field and are not central galaxies in cluster-sized
haloes where the effects of cooling flows is expected to be most
pronounced.}

\begin{figure*}
\begin{minipage}{172mm}
\begin{center}
$\begin{array}{cc}
\includegraphics[width=0.5\textwidth]{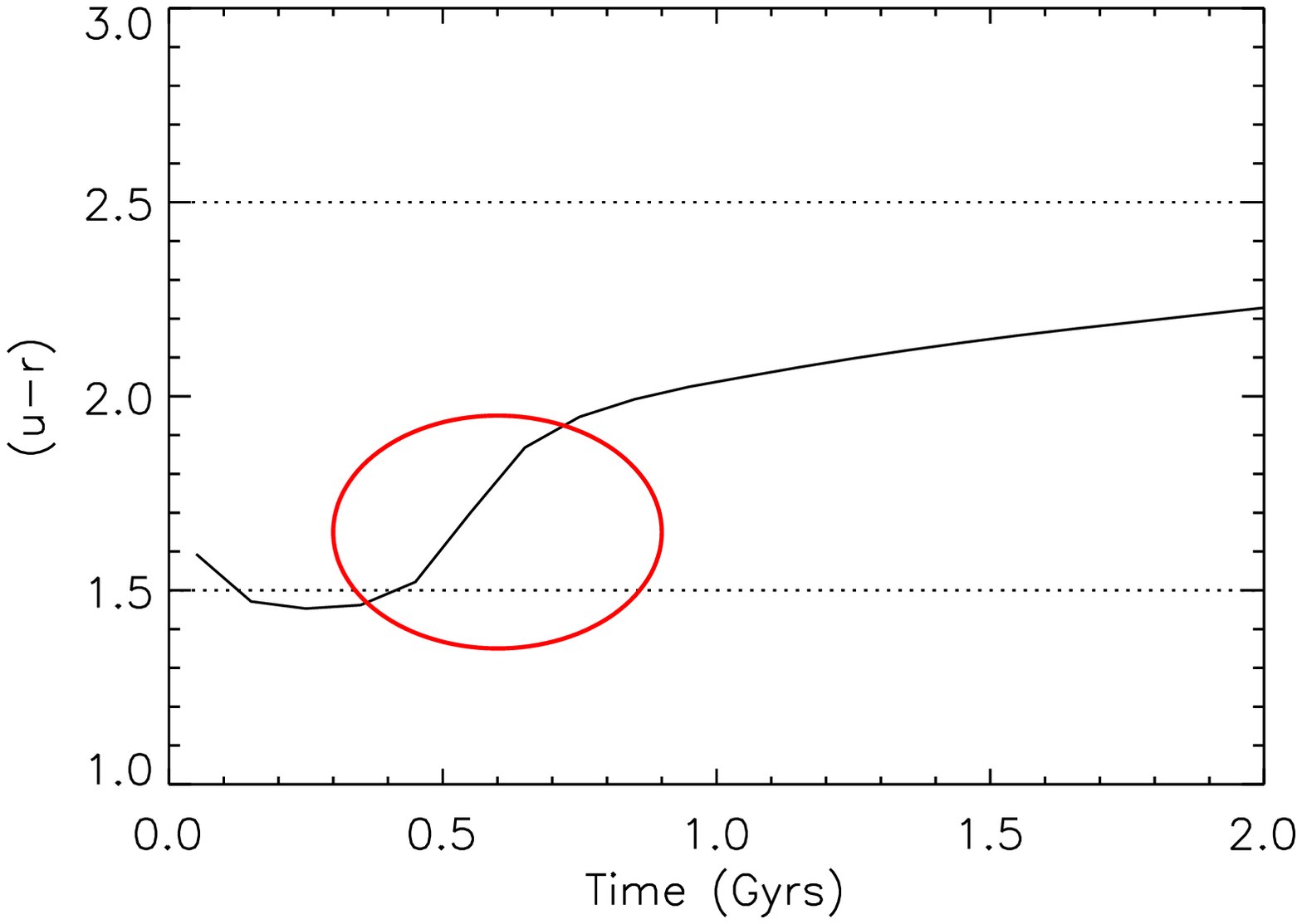} &
\includegraphics[width=0.5\textwidth]{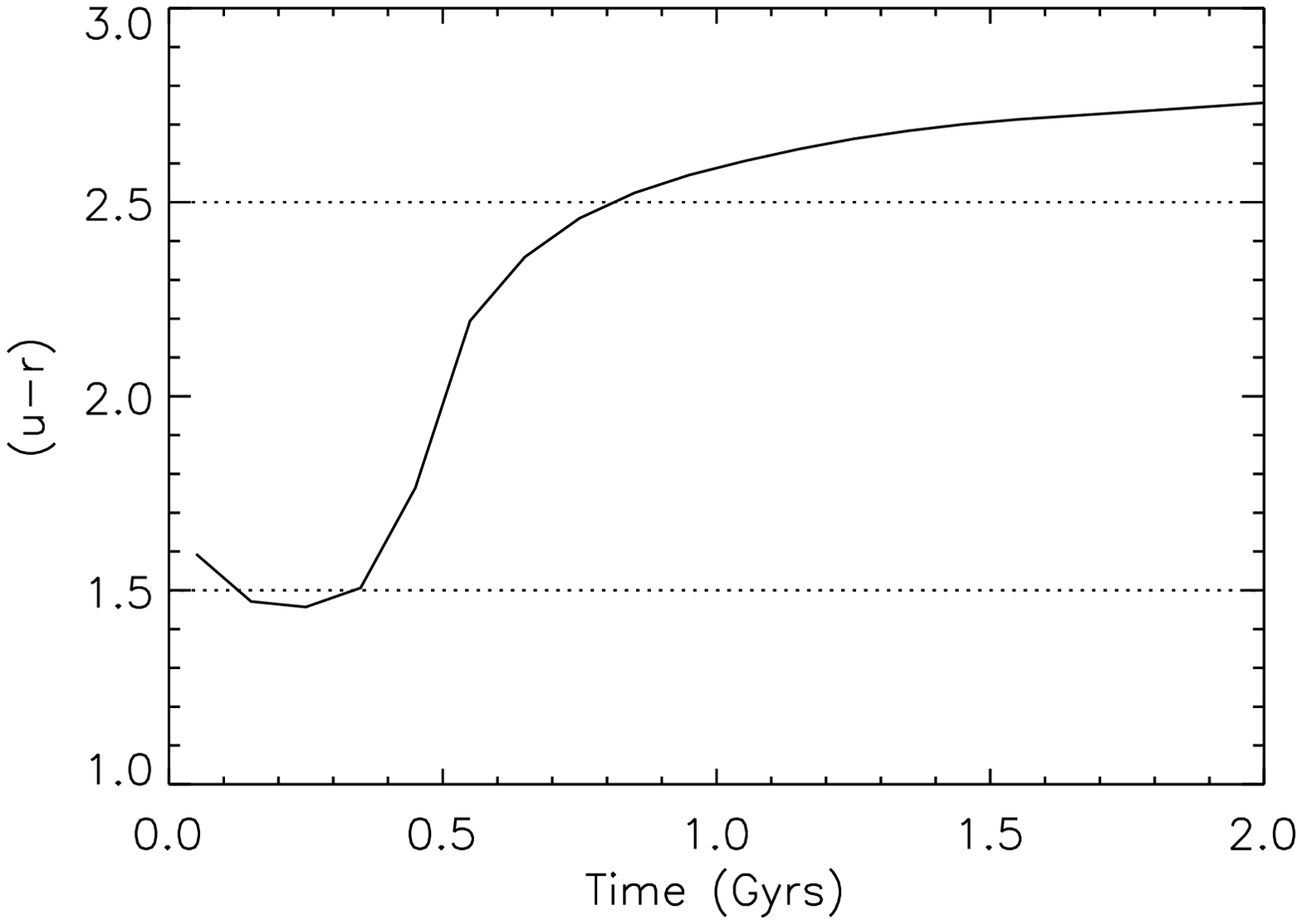}
\end{array}$
\caption{Two examples of feedback scenarios. The removal of gas
from the reservoir is most efficient around $t_p=0.5$ Gyrs, the
point at which the AGN feedback reaches its peak (see Section 3).
In the left-hand panel, we show a weak feedback scenario with
parameters ($f_0$,$\tau$) = ($10^{-3}$, 0.07 Gyrs). The reddening
induced in this scenario (highlighted by the red ellipse) is not
fast enough to achieve the migration times in the S07 study
because the galaxy is still in the `green valley' after $\sim$2
Gyrs. In contrast, the right-hand panel shows a scenario which
reproduces the fast migration times observed in S07, with
parameters ($f_0$,$\tau$) = (0.005, 0.13 Gyrs). The galaxy
transits the gap between the blue cloud and the red sequence
within around a Gyr. Note, when comparing to Figure 1, that the
time axes are shown only to 2 Gyrs in this figure.}
\label{fig:nfback_example}
\end{center}
\end{minipage}
\end{figure*}


\section{Application to a typical star-forming early-type galaxy in the blue cloud}
We proceed by exploring feedback scenarios where a \emph{typical}
star-forming {\color{black}early-type galaxy} in the blue cloud
transits the gap between the blue cloud and the red sequence, in a
manner consistent with the results of S07 and S09. Since the S07
results indicate that the mass fraction in young stars remains
virtually constant through the reddening sequence, these
{\color{black}early-type galaxies} are the likely progenitors of
the galaxies that are transiting via the green valley through to
the red sequence.

\begin{figure*}
\begin{minipage}{172mm}
\begin{center}
$\begin{array}{cc}
\includegraphics[width=0.5\textwidth]{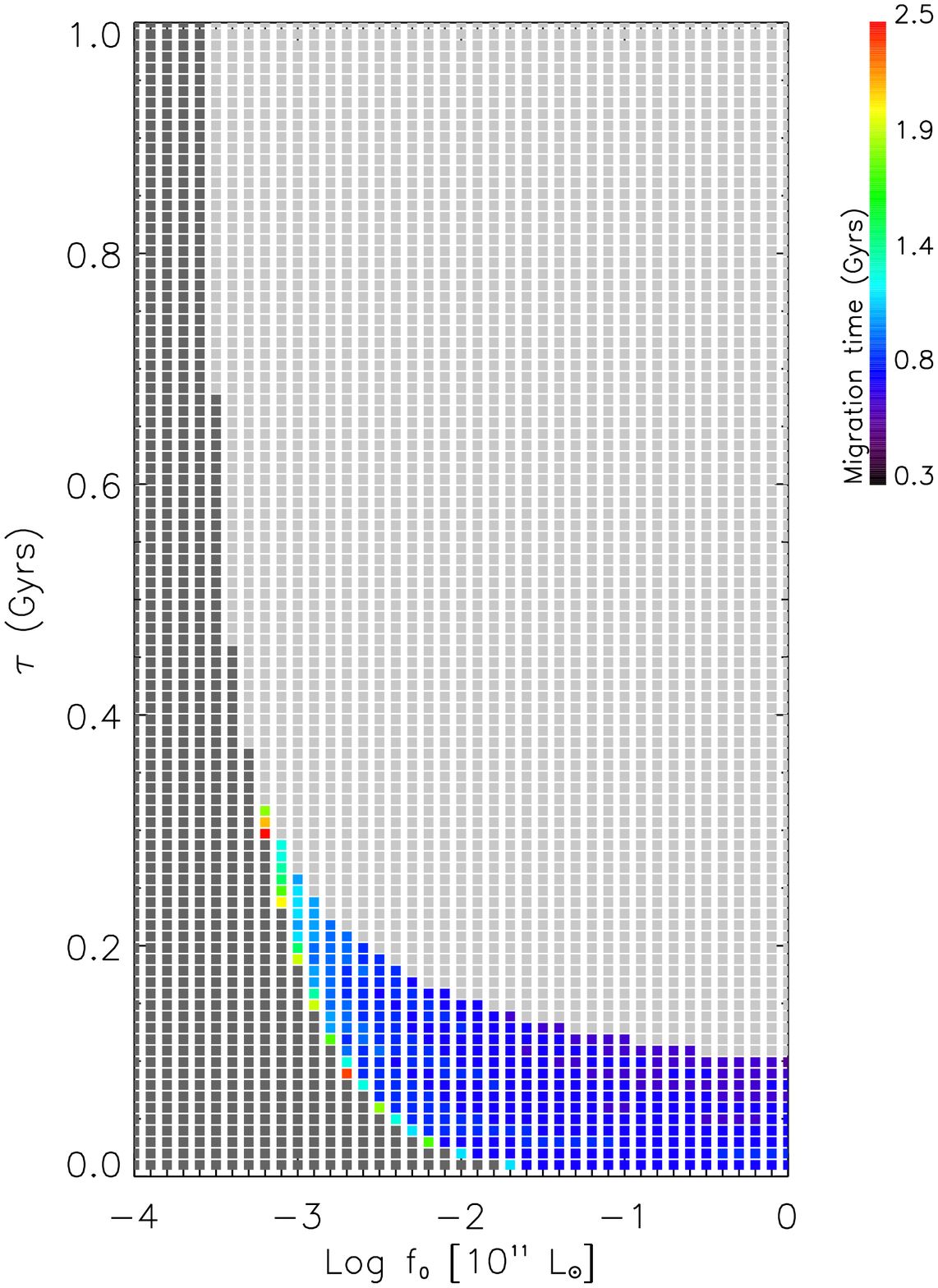} &
\includegraphics[width=0.5\textwidth]{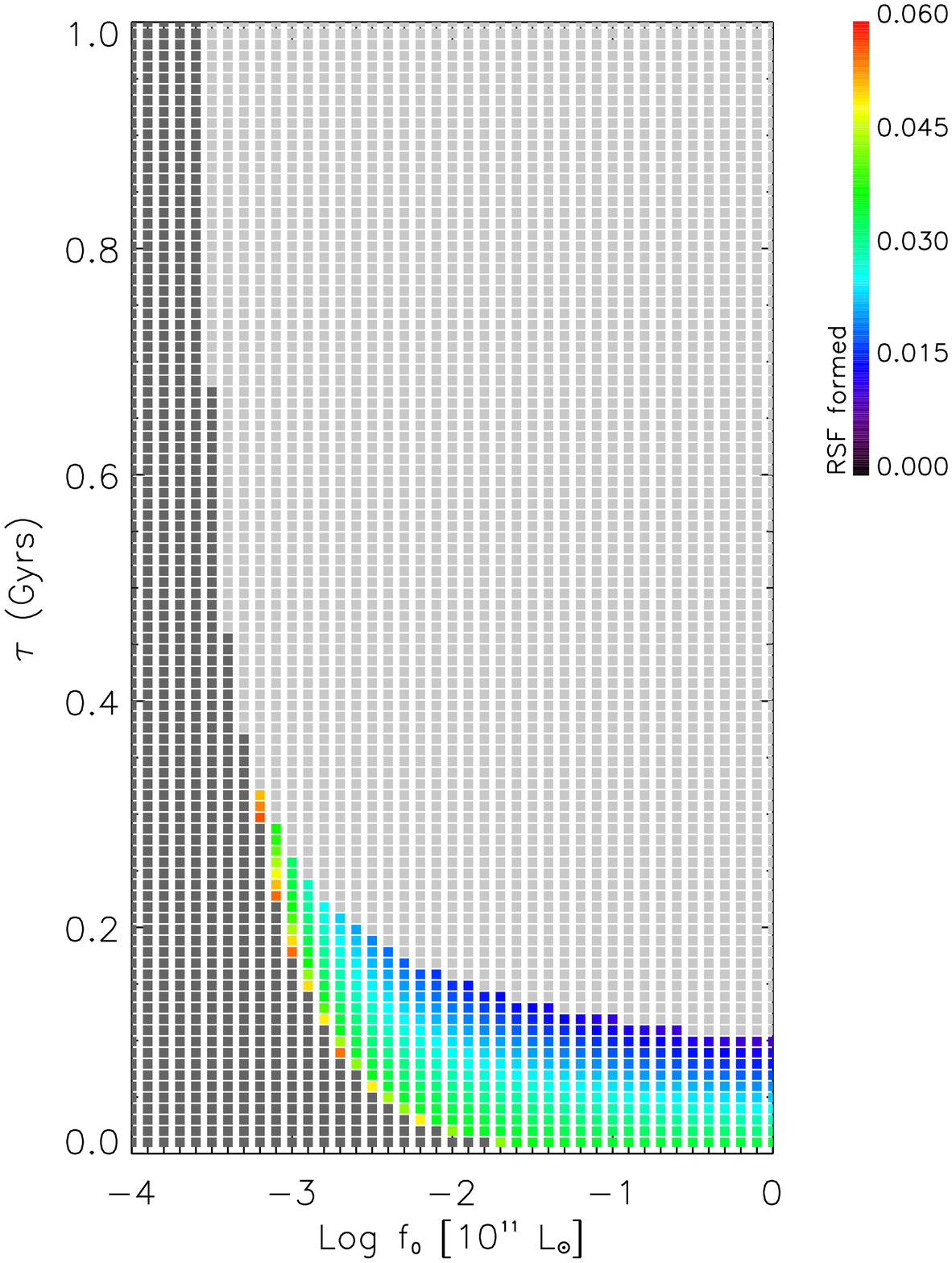}
\end{array}$
\caption{LEFT: The migration times from a set of feedback
scenarios where $10^{-4} < f_0 < 1$ and $0.01 < \tau < 1$ Gyrs.
The dark grey regions of the plot are not allowed because the
feedback is too weak - models in this region have migration times
in excess of $\sim2.5$ Gyrs. The light grey regions on the right
are not allowed because the feedback is too strong. Note that the
scenarios shown coloured in this plot bracket the S07 migration
times (0.5-2 Gyrs). RIGHT: The {\color{black} mass fractions in
young stars} for the scenarios shown on the left.}
\label{fig:nresults}
\end{center}
\end{minipage}
\end{figure*}

To set up our model in terms of a typical star-forming
{\color{black}early-type galaxy}, we require typical values of the
parameters that determine the feedback in Eqn 2. i.e. the mass
($M$) and radius ($R$) of the system. The median stellar mass of
star-forming {\color{black}early-types} in S07 is $\sim 5 \times
10^{10}$ M$_{\odot}$. We use the petrosian radius in the $r$-band,
given by the SDSS \texttt{petrorad} parameter, as a measure of the
galaxy radius. The median value of \texttt{petrorad} for the
{\color{black}early-types} in S07 is $\sim 8$ kpc. We also require
a value for the reference luminosity ($L_B$), for which we use the
median AGN luminosity in the `Seyfert' region of the S07
{\color{black}early-types}. The bolometric luminosities of AGN can
be calculated from the [OIII] $\lambda$5007 emission line
luminosities - $L_B/L_{[OIII]} \sim 3500$ with a scatter of 0.38
dex \citep[see][and references therein]{Heckman2004}. The typical
Seyfert region [OIII] luminosity in the S07
{\color{black}early-type} sample is $\sim 10^{7.5}$ L$_{\odot}$.
Thus, in what follows, we take the parameters $M$, $R$ and $L_B$
to be $5 \times 10^{10}$M$_{\odot}$, 8 kpc and $10^{11}$
L$_{\odot}$ respectively.

In a similar vein to Section 2, we construct scenarios where the
object begins its evolutionary track around $(u-r)=1.5$, which
represents the mean colour of the bluest 25\% of the star-forming
{\color{black}early-type galaxy} population. The transit times to
the `bottom' of the red sequence ($u-r \sim 2.5$) are in the range
0.5-2 Gyrs (the typical value is around a Gyr, see Figure 10 of
S07). Thus our goal is to search for solutions where objects
migrate between $u-r=1.5$ and $u-r=2.5$ within these transit
times.

We begin by showing the general impact of feedback on the colour
evolution of a model {\color{black}early-type galaxy}. The
left-hand panel of Figure \ref{fig:nfback_example} shows a
scenario where ($f_0$,$\tau$) = ($10^{-3}$, 0.07 Gyrs). The
removal of {\color{black}cold} gas from the reservoir is most
efficient around $t_p$, the point at which the feedback reaches
its peak. Recall that, following the results of S07, we assume
$t_p=0.5$ Gyrs in our model. Around $t_p$ the gas fraction in the
system experiences its sharpest decline, which induces a faster
reddening in the $(u-r)$ colour than can be achieved through star
formation evolution alone (indicated using the red ellipse). It is
evident, however, that in this particular scenario, the feedback
is not strong enough to produce the fast colour evolution observed
in S07, since the galaxy still exhibits `green valley' colours
($u-r \sim 2.2$) after 2 Gyrs of evolution.

We proceed by exploring the ($f_0$,$\tau$) parameter space to
search for scenarios where the $(u-r)$ reddening rate is
consistent with the migration times in S07. We study scenarios
where $10^{-4} < f_0 < 1$ and $0.01 < \tau < 1$ Gyrs. The
right-hand panel of Figure \ref{fig:nfback_example} shows a
scenario where ($f_0$,$\tau$) = ($0.005$, 0.12 Gyrs), which
reproduces the observed transit times reported by S07. The
migration time in this model is $\sim$0.8 Gyrs and the mass
fraction in young stars is $\sim$3\%. The left-hand panel of
Figure \ref{fig:nresults} presents a summary of the migration
times for the scenarios discussed above, while the right-hand
panel shows the {\color{black}mass fractions in young stars
forming in each model.} The dark grey regions of the plot are not
allowed because the feedback is too weak - models in this region
have migration times in excess of $\sim2.5$ Gyrs. Note that the
scenarios shown coloured in Figure \ref{fig:nresults} bracket the
S07 migration times (0.5-2 Gyrs). The light grey regions on the
right are not allowed because the feedback is too strong. In these
scenarios, the {\color{black}cold} gas mass is depleted so quickly
that the galaxy never gets a chance to reach $u-r \sim 1.5$ in the
first place.

We find that a part of the parameter space does satisfy the
migration times (shown colour-coded) observed in S07. These
scenarios typically have feedback timescales $\lesssim 0.2$ Gyrs
(left-hand panel) and produce {\color{black} mass fractions in
young stars} ranging from less than a percent to $\sim 5$ percent
(right-hand panel). Not unexpectedly, the $\tau$ and $f_0$ values
in these acceptable scenarios are to some extent degenerate in the
expected way - longer AGN timescales require lower coupling
efficiencies and vice versa. We find that, for these models, the
residual {\color{black}cold} gas fractions are $\lesssim 0.6$\%,
with the gas reservoirs already depleted by the time the galaxy
arrives in the green valley, in good agreement with the results of
S09. It is interesting to note that the coupling is relatively
weak while the galaxy is in the blue cloud, which may be
consistent with an apparent lack of high-luminosity AGN in blue
{\color{black}early-types} (Schawinski et al. al 2009b).

Our results suggest that, to achieve the migration times observed
in S07, the original reservoir of {\color{black}cold} gas in the
star-forming {\color{black}early-types} must be almost completely
evacuated by the time the galaxy approaches the red sequence.
Furthermore, since the {\color{black} mass fractions in young
stars} are less than $\sim 5$ percent, more than half of the
available fuel for star formation is likely to be lost to the
inter-galactic medium. Note that, since the measured
{\color{black} young-star} fractions in the blue S07
{\color{black}early-types} are typically a few percent, scenarios
which produce very small {\color{black} mass} fractions (e.g. less
than a percent) are unlikely. If we restrict ourselves further to
the feedback timescales that are consistent with expected AGN duty
cycles \citep[a few hundred Myrs,
e.g.][]{Haehnelt1998,Martini2001,Mathur2001,Shabala2008}, then
only a few tenths of a percent ($10^{-3}$ to $10^{-2}$) of the
luminosity of a typical Seyfert AGN must couple to the
{\color{black}cold} gas reservoir to produce the observed
migration times. Such scenarios produce mass fractions in young
stars around 2-4\% and leave remnants with {\color{black}cold} gas
fractions $\lesssim 0.6$\%, in good agreement with the data.

Finally, it is worth comparing the range of $f_0$ values derived
in this study with similar estimates obtained using other methods.
\citet{Ciotti2010} find that (radiation-driven) outflows only
require coupling efficiencies of $10^{-3}$ to $10^{-2}$ between
the AGN luminosity and the ambient {\color{black}cold} gas to
completely remove the gaseous component of the galaxy. Radio
source modelling of the interaction between jets and their
environments \cite[e.g.][]{DeYoung1993,Sutherland2007} suggest
that values of $10^{-3}$ to $10^{-1}$ are possible.
Observationally, studies of X-ray cavities and associated radio
cocoons have revealed coupling efficiencies between $10^{-4}$ and
$10^{-2}$ \citep{Birzan2004}. {\color{black}The coupling strengths
derived here (a few tenths of a percent) therefore appear
reasonably consistent with the aforementioned studies, which have
derived coupling efficiencies using independent methods.}


\section{Summary}
A growing body of recent observational evidence now indicates that
star formation at late epochs ($z<1$) adds a significant minority
of the stellar mass ($10-15$\%) in massive early-type galaxies
\citep[e.g.][]{Kaviraj2008b}. {\color{black}While energetic
arguments have made a compelling theoretical case for AGN feedback
to regulate star formation in galaxy formation models,
observational constraints on this putative process in the nearby
{\color{black}early-type galaxy population} has remained limited
but are desirable.}

In this paper, we have presented a strong plausibility argument
for AGN feedback to play a significant role in the evolution of
{\color{black}star-forming early-types} at late epochs. Previous
work in S07 has shown that star-forming {\color{black}early-types}
in the blue cloud show evidence for rapid migration to the red
sequence, typically within a Gyr, passing through several phases
of AGN activity as they move from the blue cloud ($u-r \sim 1.5$)
to the red sequence. A standard `BPT' analysis, using optical
emission line ratios, indicates that {\color{black}early-types}
classified as `star-forming' are the bluest, with those classified
as `composites', `Seyferts', `LINER' and 'quiescent' becoming
progressively redder in that order. This is accompanied by a
precipitous decrease in the {\color{black}cold} gas mass in the
system of an order of magnitude between the star-forming and
Seyfert phases (S09).

Using recent results which indicate that star formation in
{\color{black}early-types} can be described by the empirical
{\color{black}Schmidt-Kennicutt} law, we have studied whether
natural depletion of the {\color{black}cold} gas reservoir in
star-forming {\color{black}early-type galaxies}, purely through
{\color{black}Schmidt-Kennicutt}-driven star formation, can
produce the rapid colour migration observed in S07. We have shown
that this colour migration is a few factors too slow, compared to
what is observed by S07, essentially because the gas depletion
rate is not adequately fast. It is therefore reasonable to suggest
that, to achieve the observed reddening rate, an additional
mechanism is required to accelerate the depletion of
{\color{black}cold} gas and induce a faster transition from the
blue cloud to the red sequence. The coincidence of AGN activity
and the rapid observed colour transition strongly suggests that
the AGN may play a significant role in driving this gas depletion
and the transit of galaxies from the blue cloud to the red
sequence.

To explore the broad characteristics of this AGN-driven feedback
we have developed a simple phenomenological model in which a
fraction of the bolometric luminosity of the AGN couples to the
{\color{black}cold} gas reservoir and removes some of the gas
mass, while the remaining gas continues to produce stars according
to the {\color{black}Schmidt-Kennicutt} law. The impact of this
feedback is to accelerate the rate of gas depletion, which induces
a faster colour transition that is consistent with the results of
S07. Our results suggest that a few tenths of a percent of the
luminosity of a Seyfert AGN ($\sim 10^{11}$ L$_{\odot}$) must
couple to the {\color{black}cold} gas reservoir of a typical
star-forming {\color{black}early-type galaxy} over a duty cycle (a
few hundred Myrs) to induce a colour transition that is consistent
with the findings of S07 and S09. Such scenarios lead to mass
fractions in young stars of a few percent, with residual
{\color{black}cold} gas fractions of less than $\sim$ 0.6\%, both
consistent with the measurements of {\color{black} mass fractions
of young stars} in blue {\color{black}early-types} and residual
{\color{black}cold} gas masses in Seyfert
{\color{black}early-types}. As we discuss in Section 4 above, the
coupling efficiencies derived here are consistent with
independently derived values in the literature.

We conclude by connecting the results in this paper to recent work
on the evolution of {\color{black}early-type galaxies} in the
local Universe. As noted in the introduction, both theoretical and
observational arguments indicate that the {\color{black} recent
star formation} in {\color{black}early-types} is likely to be
influenced by minor merging at late epochs. This is supported by
evidence for kinematical decoupling between the (ionised) gas and
stars and the fact that the gas mass shows no correlation with the
stellar mass of the galaxy, irrespective of the local environment,
both indicating that the gas is, at least in part, external in
origin. The {\color{black}cold} gas injected by gas-rich infalling
satellites is likely to trigger low-level star formation which
moves the spheroid temporarily to the blue cloud. This is followed
by the fuelling of the central black hole and AGN activity, which
reaches a peak around 0.5 Gyrs after the onset of star formation.
In the time delay between the onset of star formation and the peak
of the AGN activity, the induced star formation adds a few percent
(or less) to the stellar mass of the original spheroid. The rise
of the AGN then acts to rapidly quench the star formation and
restores the spheroid to the red sequence over a short timescale
($\sim$ 1 Gyr).


\nocite{Schawinski2009b}


\section*{Acknowledgements}
We are grateful to the anonymous referee for a careful review
which helped clarify several sections of the paper. SK
acknowledges a Research Fellowship from the Royal Commission for
the Exhibition of 1851, an Imperial College Junior Research
Fellowship, a Senior Research Fellowship from Worcester College,
Oxford and support from the BIPAC institute at Oxford. It is a
pleasure to thank Hagai Netzer and Martin Bureau for stimulating
discussions. Support for KS was provided by NASA through Einstein
Postdoctoral Fellowship grant number PF9-00069 issued by the
Chandra X-ray Observatory Center, which is operated by the
Smithsonian Astrophysical Observatory for and on behalf of NASA
under contract NAS8-03060. SS acknowledges a research fellowship
from New College, Oxford.


\nocite{Pipino2009}


\bibliographystyle{mn2e}
\bibliography{references}


\end{document}